\RequirePackage{fix-cm}
\RequirePackage{fixltx2e}
\documentclass[reqno]{amsproc}
\usepackage{amsthm,amssymb}
\usepackage{mathtools}

\usepackage[citation-order,nobysame]{amsrefs}
\usepackage{xyzbib}

\usepackage{pstricks,pst-node,pst-plot}
\usepackage{hyperref}

\mathtoolsset{showonlyrefs}

\numberwithin{equation}{section}
\let\cite=\cites
\newcommand{\rmd}{\mathrm{d}}
\newcommand{\rmi}{\mathrm{i}}

\newcommand{\ac}{\alpha_\mathrm{c}}
\newcommand{\yc}{y_\text{c}}
\newcommand{\Rc}{R_\text{c}}
\newcommand{\Qc}{Q_\text{c}}

\newtheorem{theorem}{Theorem}[section]
\newtheorem{proposition}[theorem]{Proposition}
\newtheorem{remark}[theorem]{Remark}
\begin{document}

\title{Thermodynamics of the six-vertex model in an L-shaped domain}

\author{Filippo Colomo}
\address{INFN, Sezione di Firenze,
Via G. Sansone 1, Sesto Fiorentino (FI), 50019, Italy}
\email{colomo@fi.infn.it}

\author{Andrei G. Pronko}
\address{Steklov Mathematical Institute, Fontanka 27,
St. Petersburg, 191023, Russia}
\email{agp@pdmi.ras.ru}

\begin{abstract}
We consider the six-vertex model in an L-shaped domain of the square lattice,
with domain wall boundary conditions.  For free-fermion vertex weights the
partition function can be expressed in terms of some Hankel determinant, or
equivalently as a Coulomb gas with discrete measure and a non-polynomial
potential with two hard walls. We use Coulomb gas methods to study the
partition function in the thermodynamic limit.  We obtain the free energy
of the six-vertex model as a function of the parameters describing the
geometry of the scaled L-shaped domain. Under variations of these
parameters the system undergoes a third-order phase transition. The result
can also be considered in the context of dimer models, for the perfect
matchings of the Aztec diamond graph with a cut-off corner.
\end{abstract}

\maketitle
\section{Introduction}

It is commonly known that macroscopic quantities (such as the free
energy) of dimer coverings and random tilings on regular lattices may
depend on boundary conditions \cite{GCZ-80,EKLP-92}.  This feature is
related to phase separation and the emergence of a limit shape
\cite{JPS-98,CLP-98,KO-05,KOS-06,KO-06}, see \cite{K-09} for a review.
The same phenomena can be observed in the six-vertex model
\cite{KZj-00,RP-06,CP-09}.  In relation with these phenomena, an
interesting question concerns the stability of the observed bulk
properties against various deformations of the shape of the considered
finite region, while preserving the particular boundary conditions
(e.g., the staircase shape for the boundary of the Aztec diamond in
the case of domino tilings).

In this paper, we address the problem by studying the
thermodynamics of the six-vertex model on a particular domain of the
square lattice, that we call an L-shaped domain. The model is closely
related to the domino tilings of the Aztec diamond with a cuf-off
corner \cite{CP-13}. We evaluate the exact analytic expression for the
free energy per site of the six-vertex model at its free fermion
point, as a function of the parameters describing the geometry of the
scaled L-shaped domain.  Under variation of these parameters, when the
boundary interferes with the phase separation curve, the system
undergoes a third-order phase transition.

Our result on the free energy of the six-vertex model in an L-shaped
domain is based on some discrete Coulomb gas representation.  Discrete
Coulomb gases already appeared in applications to tiling and dimer
models \cite{J-00,J-02}, and to the six-vertex model
\cite{Zj-00,BL-13}, and have been studied intensively in the context
of discrete orthogonal polynomials, see \cite{BKMM-07} and references
therein.  As a side result, we find that the soft-edge/hard-edge
transition may induce (or not) a third-order phase transition in the
free energy, according to the finite (or diverging) slope of the
discrete Coulomb gas potential in the vicinity of the hard wall.

\subsection{The model}

The states of the six-vertex model are configurations of arrows on the
edges of a square lattice, satisfying the \emph{ice rule}: at each vertex
the numbers of incoming and outgoing arrows are equal.  The ice rule
selects $\binom{4}{2}=6$ possible vertex configurations, hence the
name of the model.

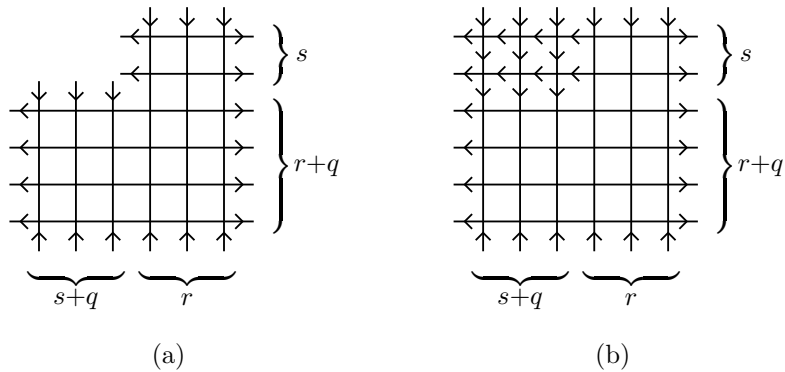
\begin{figure}
\centering

\psset{unit=14pt,linewidth=.05}
\newcommand{\arr}{\lput{:U}{\begin{pspicture}(0,0)
\psline(-.1,.2)(.1,0)(-.1,-.2) \end{pspicture}}}

\begin{pspicture}(0,-3)(21,7)

\multirput(1,1)(1,0){6}{\psline(0,0)(0,3)}
\multirput(1,1)(0,1){4}{\psline(0,0)(5,0)}
\multirput(4,4)(1,0){3}{\psline(0,0)(0,2)}
\multirput(4,5)(0,1){2}{\psline(0,0)(2,0)}
\multirput(1,0)(1,0){6}{\pcline(0,0.2)(0,1)\arr}
\multirput(0,1)(0,1){6}{\pcline(6,0)(6.8,0)\arr}
\multirput(0,1)(0,1){4}{\pcline(1,0)(.2,0)\arr}
\multirput(3,5)(0,1){2}{\pcline(1,0)(.2,0)\arr}
\multirput(1,0)(1,0){3}{\pcline(0,4.8)(0,4)\arr}
\multirput(4,0)(1,0){3}{\pcline(0,6.8)(0,6)\arr}

\rput{90}(7.5,2.5)
  {$\underbrace{\begin{pspicture}(0.2,0)(3.8,0)\end{pspicture}}$}
\rput(8.5,2.5){$r{+}q$}

\rput{0}(2,-.5)
 {$\underbrace{\begin{pspicture}(0.2,0)(2.8,0)\end{pspicture}}$}
\rput(2,-1.1){$s{+}q$}

\rput{0}(5,-.5)
  {$\underbrace{\begin{pspicture}(0.2,0)(2.8,0)\end{pspicture}}$}
\rput(5,-1.1){$r$}

\rput{90}(7.5,5.5)
  {$\underbrace{\begin{pspicture}(0.2,0)(1.8,0)\end{pspicture}}$}
\rput(8.1,5.5){$s$}

\rput[b](4.5,-3){(a)}

\rput(12,0)
{
\multirput(1,0)(1,0){6}{\pcline(0,0.2)(0,1)\arr}
\multirput(0,1)(0,1){6}{\pcline(6,0)(6.8,0)\arr}
\multirput(0,1)(0,1){6}{\pcline(1,0)(.2,0)\arr}
\multirput(1,0)(1,0){6}{\pcline(0,6.8)(0,6)\arr}
\multirput(4,1)(1,0){3}{\psline(0,0)(0,5)}
\multirput(1,1)(1,0){3}{\psline(0,0)(0,3)}
\multirput(1,1)(0,1){4}{\psline(0,0)(5,0)}
\multirput(1,5)(0,1){2}{\psline(3,0)(5,0)}
\multirput(1,6)(1,0){3}{\pcline(1,0)(0,0)\arr}
\multirput(1,5)(1,0){3}{\pcline(1,0)(0,0)\arr}
\multirput(1,5)(1,0){3}{\pcline(0,1)(0,0)\arr}
\multirput(1,4)(1,0){3}{\pcline(0,1)(0,0)\arr}

\rput{90}(7.5,2.5)
  {$\underbrace{\begin{pspicture}(0.2,0)(3.8,0)\end{pspicture}}$}
\rput(8.5,2.5){$r{+}q$}

\rput{0}(2,-.5)
 {$\underbrace{\begin{pspicture}(0.2,0)(2.8,0)\end{pspicture}}$}
\rput(2,-1.1){$s{+}q$}

\rput{0}(5,-.5)
  {$\underbrace{\begin{pspicture}(0.2,0)(2.8,0)\end{pspicture}}$}
\rput(5,-1.1){$r$}

\rput{90}(7.5,5.5)
  {$\underbrace{\begin{pspicture}(0.2,0)(1.8,0)\end{pspicture}}$}
\rput(8.1,5.5){$s$}

\rput[b](4.5,-3){(b)}
}

\end{pspicture}
\caption{(a) The L-shaped domain with domain wall boundary
conditions; (b) The corresponding arrow configuration on the
$N\times N$ lattice.}
\label{fig-LshapedDomain}
\end{figure}

The L-shaped domain can be defined as a square domain with a
rectangular portion removed from one of the corners, see
Fig.~\ref{fig-LshapedDomain}. Specifically, the
square domain is the finite square lattice obtained from the
intersection of $N$ horizontal and $N$ vertical lines, the so-called
$N \times N$ lattice. The L-shaped domain is obtained by removing from
the top left corner a rectangular portion of the lattice, of size
$s\times (s+q)$, $s,q\in\mathbb{N}_0$, $s+q<N$. The L-shaped
domain is shown in Fig.~\ref{fig-LshapedDomain}a, where $r=N-s-q$.

We consider a specific instance of fixed boundary conditions, the
domain wall boundary conditions. In the standard arrows picture of the
six-vertex model (see, e.g., \cite{B-82}) this conditions means that
all horizontal (respectively, vertical) arrows on external edges are
outgoing (incoming). For $s=0$, one has the usual $N\times N $ lattice
with domain wall boundary conditions, introduced in \cite{K-82}.

\begin{figure}
\centering

\psset{unit=10pt,dotsep=2pt}
\newcommand{\arr}{\lput{:U}{\begin{pspicture}(0,0)
\psline(-.05,.2)(.15,0)(-.05,-.2) \end{pspicture}}}

\begin{pspicture}(0,0)(20,5)
\rput(0,0){
\psline[linestyle=dotted](0,0)(0,2)(2,2)(2,0)(0,0)
\pcline(0,1)(1,1)\arr \pcline(1,1)(2,1)\arr
\pcline(1,0)(1,1)\arr \pcline(1,1)(1,2)\arr
}
\rput(3,0){
\psline[linestyle=dotted](0,0)(0,2)(2,2)(2,0)(0,0)
\pcline(1,1)(0,1)\arr \pcline(2,1)(1,1)\arr
\pcline(1,2)(1,1)\arr \pcline(1,1)(1,0)\arr
}
\rput(6,0){
\psline[linestyle=dotted](0,0)(0,2)(2,2)(2,0)(0,0)
\pcline(0,1)(1,1)\arr \pcline(1,1)(2,1)\arr
\pcline(1,2)(1,1)\arr \pcline(1,1)(1,0)\arr
}
\rput(9,0){
\psline[linestyle=dotted](0,0)(0,2)(2,2)(2,0)(0,0)
\pcline(2,1)(1,1)\arr \pcline(1,1)(0,1)\arr
\pcline(1,0)(1,1)\arr \pcline(1,1)(1,2)\arr
}
\rput(12,0){
\psline[linestyle=dotted](0,0)(0,2)(2,2)(2,0)(0,0)
\pcline(0,1)(1,1)\arr \pcline(2,1)(1,1)\arr
\pcline(1,1)(1,0)\arr \pcline(1,1)(1,2)\arr
}
\rput(16.5,0){
\psline[linestyle=dotted](0,0)(0,2)(2,2)(2,0)(0,0)
\pcline(1,1)(2,1)\arr \pcline(1,1)(0,1)\arr
\pcline(1,2)(1,1)\arr \pcline(1,0)(1,1)\arr
}
\rput(0,3){
\psline[linestyle=dotted](0,0)(0,2)(2,2)(2,0)(0,0)
\psline(0,0)(2,2)\psline(1,1)(0,2)
}
\rput(3,3){
\psline[linestyle=dotted](0,0)(0,2)(2,2)(2,0)(0,0)
\psline(0,0)(2,2)\psline(1,1)(2,0)
}
\rput(6,3){
\psline[linestyle=dotted](0,0)(0,2)(2,2)(2,0)(0,0)
\psline(0,2)(2,0)\psline(1,1)(0,0)
}
\rput(9,3){
\psline[linestyle=dotted](0,0)(0,2)(2,2)(2,0)(0,0)
\psline(0,2)(2,0)\psline(1,1)(2,2)
}
\rput(12,3){
\psline[linestyle=dotted](0,0)(0,2)(2,2)(2,0)(0,0)
\psline(0,0)(2,2)\psline(2,0)(0,2)
}
\rput(15,3){
\psline[linestyle=dotted](0,0)(0,2)(2,2)(2,0)(0,0)
\psline(0,0)(2,2)
}
\rput(18,3){
\psline[linestyle=dotted](0,0)(0,2)(2,2)(2,0)(0,0)
\psline(2,0)(0,2)
}
\rput(17.5,2.5){$\underbrace{\begin{pspicture}(-.2,0)(5.2,0)\end{pspicture}}$}
\end{pspicture}

\caption{Small patches of dominoes (first row)
and vertex configurations of the six-vertex model (second row).}
\label{fig-DominoPatches}
\end{figure}
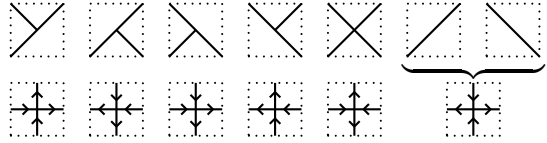

The Boltzmann weights of the six-vertex model, $w_i$, $i=1,\dots,6$,
enumerated in the standard order, see Fig.~\ref{fig-DominoPatches},
second row, are chosen to satisfy
\begin{equation}\label{FF}
w_1w_2+w_3w_4=w_5w_6,
\end{equation}
that is called the free-fermion condition.

We denote the partition function of the six-vertex model in an
L-shaped domain with domain wall boundary conditions by $Z_{r,s,q}$;
for the special case $s=0$, we use the standard notation $Z_N$.  For
generic Boltzmann weights, the partition function $Z_N$ has been evaluated
in determinantal form in \cite{I-87,ICK-92}. For weights satisfying
the condition \eqref{FF}, $Z_N$ has an extremely concise form:
\begin{equation}\label{ZN}
Z_N=w_5^{\frac{N(N-1)}{2}}w_6^{\frac{N(N+1)}{2}}.
\end{equation}
For a proof, see \cite{EKLP-92} in the context of enumerative
combinatorics, or \cite{BPZ-02} in the context of
integrable models.

For generic $s$, the partition function $Z_{r,s,q}$ can be expressed
in terms of certain non local correlation function of the model on the
$N\times N$ lattice. This function is called the emptiness formation
probability, and denoted here as $F_{r,s,q}$.  It describes the
probability that all vertices in the top left $s\times (s+q)$
rectangle of the $N\times N$ lattice are of type 2. The corresponding
configuration is shown in Fig.~\ref{fig-LshapedDomain}b.  Comparing
the two pictures of Fig.~\ref{fig-LshapedDomain}, it is clear that
\begin{equation}\label{Zrsq}
Z_{r,s,q}=\frac{Z_N}{w_2^{s(s+q)}}F_{r,s,q}.
\end{equation}
Thus the partition function of the six-vertex model on the L-shaped domain
is essentially given by the emptiness formation probability of the
model on the $N\times N$ lattice.

The above construction can be translated in the language of dimer models,
using the well known correspondence between the six-vertex model with
domain wall boundary conditions and the domino tilings of the Aztec
diamond \cite{EKLP-92}.

\begin{figure}
\centering

\psset{unit=14pt,linewidth=.05}
\newcommand{\arr}{\lput{:U}{\begin{pspicture}(0,0)
\psline(-.1,.2)(.1,0)(-.1,-.2) \end{pspicture}}}

\begin{pspicture}(0,-1.5)(19,7)

\multirput(1,0)(1,0){6}{\psline(-.5,.5)(0,0)(.5,.5)}
\multirput(7,1)(0,1){6}{\psline(-.5,.5)(0,0)(-.5,-.5)}
\multirput(1,7)(1,0){6}{\psline(-.5,-.5)(0,0)(.5,-.5)}
\multirput(0,1)(0,1){6}{\psline(.5,-.5)(0,0)(.5,.5)}

\psline(2.5,4.5)(3.5,5.5)

\psline(0.5,5.5)(1.5,6.5)
\psline(0.5,4.5)(2.5,6.5)
\psline(1.5,4.5)(3.5,6.5)
\psline(1,6)(1.5,5.5)
\psline(1,5)(1.5,4.5)
\psline(2,5)(2.5,4.5)
\psline(2,6)(2.5,5.5)
\psline(3,6)(3.5,5.5)
\psline(3,5)(3.5,4.5)

\multirput(.5,.5)(1,0){7}{\psline[linestyle=dotted,dotsep=.2,linewidth=.02](0,0)(0,6)}
\multirput(.5,.5)(0,1){7}{\psline[linestyle=dotted,dotsep=.2,linewidth=.02](0,0)(6,0)}


\rput[b](3.5,-1.5){(a)}

\rput(12,0){
\multirput(1,0)(1,0){6}{\psline(-.5,.5)(0,0)(.5,.5)}
\multirput(7,1)(0,1){6}{\psline(-.5,.5)(0,0)(-.5,-.5)}
\multirput(4,7)(1,0){3}{\psline(-.5,-.5)(0,0)(.5,-.5)}
\multirput(0,1)(0,1){4}{\psline(.5,-.5)(0,0)(.5,.5)}

\multirput(1,4)(1,0){3}{\psline(-.5,.5)(0,1)(.5,.5)}
\multirput(3,5)(0,1){2}{\psline(.5,-.5)(0,0)(.5,.5)}



\multirput(0,0)(1,0){6}{
\multirput(0,0)(0,1){4}{\psline[linewidth=.01](1,.5)(.5,1)(1,1.5)(1.5,1)(1,.5)}
}
\multirput(3,4)(1,0){3}{
\multirput(0,0)(0,1){2}{\psline[linewidth=.01](1,.5)(.5,1)(1,1.5)(1.5,1)(1,.5)}
}

\multirput(0,0)(0,1){5}{
\multirput(0,0)(1,0){6}{\psdot[dotsize=.15](1,.5)}}
\multirput(0,0)(1,0){7}{
\multirput(0,0)(0,1){4}{\psdot[dotsize=.15,dotstyle=o](.5,1)}}

\multirput(3,5)(0,1){2}{
\multirput(0,0)(1,0){3}{\psdot[dotsize=.15](1,.5)}}
\multirput(3,4)(1,0){4}{
\multirput(0,0)(0,1){2}{\psdot[dotsize=.15,dotstyle=o](.5,1)}}

\rput[b](3.5,-1.5){(b)}
}

\end{pspicture}

\caption{(a) Aztec diamond with a frozen region of NE-SW dominoes;
(b) Aztec diamond with a cut-off corner and the corresponding
modified Aztec diamond graph.}
\label{fig-AztecDiamonds}
\end{figure}
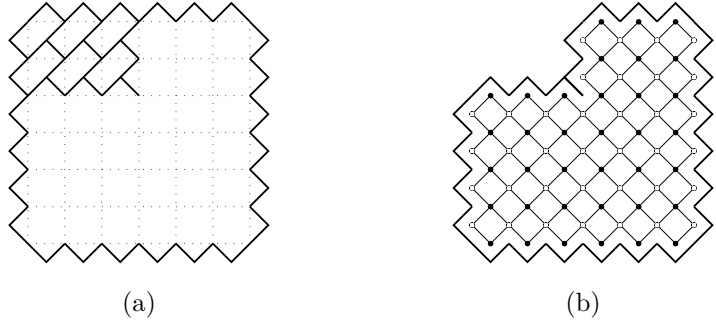

Recall that \emph{domino tilings} are coverings of a square lattice
with rectangles of size $1\times 2$ and $2\times 1$.  Domino tilings
can be formulated in terms of the six-vertex model on a square lattice
by mapping elementary patches of domino tilings to arrow
configurations as shown in Fig.~\ref{fig-DominoPatches}.  The case of
the six-vertex model on an $N\times N$ square lattice with domain wall
boundary conditions corresponds to the Aztec diamond of order $N$, see
Fig.~\ref{fig-AztecDiamonds}.

The plain enumeration of domino tilings of the Aztec diamond
corresponds to the choice of the Boltzmann weights of the six-vertex
model $w_1=\dots=w_5=1$ and $w_6=2$.  In counting of domino tilings,
one may also consider weighted enumerations, with a \emph{bias}
parameter $\alpha\in[0,1]$, describing the asymmetry between the two
possible orientations of dominoes.  Following paper \cite{JPS-98}, we
assign to the NE-SW dominoes the weight $\sqrt{2(1-\alpha)}$, and to
the NW-SE the weight $\sqrt{2\alpha}$. To establish the correspondence
with the six-vertex model, it is useful to parameterize the Boltzmann
weights as follows:
\begin{equation}\label{weights}
w_1=w_2=\sqrt{\rho(1-\alpha)},\qquad
w_3=w_4=\sqrt{\rho\alpha},\qquad
w_5=1,\qquad
w_6=\rho.
\end{equation}
Then the weighted enumeration of dominoes corresponds to setting $\rho=2$.

In the same spirit, the six-vertex model in an L-shaped domain can be
related to the domino tilings of some suitable modification of the
Aztec diamond.
Using the correspondence of vertices of type 2 with a
particular patch of domino tilings, see Fig.~\ref{fig-DominoPatches},
it is clear that emptiness formation probability is equivalent to the
probability of observing the domino configuration shown in
Fig.~\ref{fig-AztecDiamonds}a.  Removing the frozen region of NE-SW
dominoes from the Aztec diamond, one obtain the Aztec diamond with a
cut-off corner shown in Fig.~\ref{fig-AztecDiamonds}b. The same
pictures also shows the resulting graph for dimer coverings (modified
Aztec diamond graph).

\subsection{Discrete Coulomb gas representation}

As discussed above, the partition function of the six-vertex model on
an L-shaped domain is essentially given by the emptiness formation
probability. For generic values of Boltzmann weights, the emptiness
formation probability has been computed in the form of a multiple
integral in \cite{CP-07b}. Under the free-fermion condition this
representation reduces to an Hankel determinant \cite{P-13}.

\begin{proposition}\label{prop1}
The emptiness formation probability of the six-vertex model on the
$N\times N$ lattice with domain wall boundary condition, with the
Boltzmann weights \eqref{weights} admits the following
representation:
\begin{equation}\label{newEFP}
F_{r,s,q}=\frac{(q!)^s}{\prod_{k=0}^{s-1} (q+k)! k!}
\frac{(1-\alpha)^{s (s+q)}}{\alpha^{s(s-1)/2}}
\det_{1\leq j,k\leq s}
\left[\sum_{m=0}^{r-1}m^{j+k-2}\binom{m+q}{m}\alpha^m\right].
\end{equation}
\end{proposition}
For a proof, see \cite{CP-07b,P-13}.

\begin{remark}\label{remark1}
An alternative representation, equivalent to \eqref{newEFP}, reads
\begin{equation}\label{Irsqdef}
F_{r,s,q}=\frac{(1-\alpha)^{s(s+q)}}{\alpha^{s(s-1)/2}}I_{r,s,q},
\end{equation}
where
\begin{equation}\label{Irsq}
I_{r,s,q}=\frac{1}{s!}\prod_{j=0}^{s-1}\frac{q!}{j!(j+q)!}
\sum_{m_1=0}^{r-1}\cdots \sum_{m_s=0}^{r-1}
\prod_{1\leq j<k \leq s} (m_j-m_k)^2
\prod_{j=1}^{s}\binom{q+m_j}{q}\alpha^{m_j}.
\end{equation}
\end{remark}
The last representation is that of a discrete Coulomb gas confined
within a finite interval.  We recognize in the discrete weight
\begin{equation}
w_q^\alpha(m)=\binom{q+m}{q}\alpha^{m}, \qquad m\in\mathbb{N}_0 ,
\end{equation}
that of Meixner polynomials, but we emphasize the condition $m<r$ for
the charge coordinates in the Coulomb gas representation \eqref{Irsq}.
This condition plays a major role in what follows.  We shall refer to
this condition as an \emph{hard wall} for the charges.

The discrete Coulomb gas formula  \eqref{Irsq} appeared previously
in the context of some random growth model studied by Johansson
\cite{J-00}.
The connections of  \eqref{Irsq} with the circular unitary ensemble,
with a multivariate generalization of hypergeometric function, and with the
$\tau$-function of the sixth Painlev\'e equation were
enlightened by Forrester and Witte \cite{FW-04}.

\subsection{Thermodynamics: the main result}

We are interested in the behaviour of the model in the thermodynamic
limit, i.e., in the limit of the large lattice sizes,
$r,s,q\to\infty$.  This is described by the free energy per site,
defined as follows.  Let $s+q=[x N]$, $s=[y N]$, with $x,y\in[0,1]$
fixed, $x\geq y$.  We set
\begin{equation}\label{fdef}
f(x,y):=-\lim_{N\to\infty}\frac{1}{N^2-s(s+q)}\log Z_{N-s-q,s,q}.
\end{equation}
Equivalently, we may study  the emptiness formation probability
$F_{r,s,q}$. We set
\begin{equation}\label{sigmadef}
\varphi(x,y):=-\lim_{N\to\infty}\frac{1}{N^2}\log F_{N-s-q,s,q}.
\end{equation}
\begin{remark}
The above limits exist; in particular, $\varphi(x,y)$ is
non-negative all over its domain of definition, $x,y\in[0,1]$,
$x\geq y$.
\end{remark}
This is part of a stronger statement proven in \cite{J-00}, see
Theorem 1.1 therein.

From relation \eqref{Zrsq} it immediately follows that the free energy
per site of the six-vertex model on the L-shaped domain is completely
determined by the knowledge of the function $\varphi(x,y)$. Indeed, one has
\begin{equation}
(1-xy)f(x,y)=f(0,0)+xy\log w_2+\varphi(x,y),\qquad f(0,0)=-\log\sqrt{\rho}.
\end{equation}
Similarly, the function $\varphi(x,y)$ determines completely the free
energy of the domino tilings (up to trivial modifications, which can
be easily inferred from \eqref{Zrsq}).

The main result of the present paper concerns the explicit form of the
function $\varphi(x,y)$.

\begin{figure}
\centering

\psset{unit=14pt}
\begin{pspicture}(-1,-2)(12,11)
\psline(10,10)(10,0)(0,0)
\psline{->}(0,10)(12,10)
\psline{->}(0,10)(0,-2)
\rput[r](-.5,-2){$y$}
\rput[B](12,10.5){$x$}
\rput{45}(5,5){\psellipse[linewidth=.05,linestyle=dotted](0,0)(5.5,4.4)}
\rput[rB](-.5,10.5){$0$}
\rput[B](10,10.5){$1$}
\rput[r](-.5,0){$1$}
\rput[B](6.1,10.5){$1{-}\alpha$}\psline[linewidth=.05](6.1,9.8)(6.1,10.2)

\psline[linewidth=.05](10,10)(5,5)\psline[linewidth=.01](0,0)(5,5)
\psline[linewidth=.05](0,10)(5,5)\psline[linewidth=.01](10,0)(5,5)

\rput(2,9.25){$\mathcal{D}_\mathrm{I}$}
\rput(6,8.25){$\mathcal{D}_\mathrm{II}$}

\rput{45}(5,5){\psellipticarc[linewidth=.1](0,0)(5.55,4.45){33}{90}}


\end{pspicture}

\caption{The Arctic ellipse (dotted), the arc $\mathcal{A}$ (bold),
and the domains $\mathcal{D}_{\mathrm{I}}$ and $\mathcal{D}_{\mathrm{II}}$.}
\label{fig-ArcticEllipse}
\end{figure}
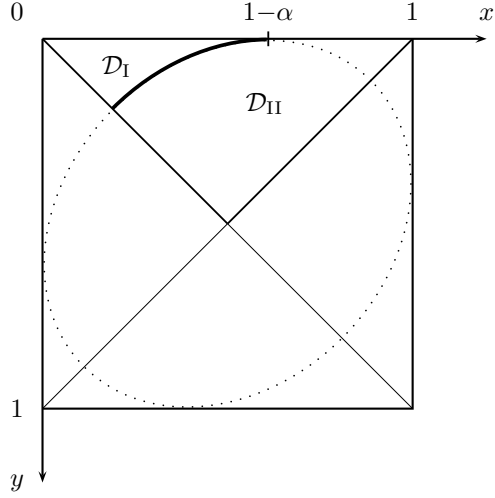

Let us consider the unit square, parameterized by $x,y\in[0,1]$.  It
follows from \eqref{sigmadef} that the domain of definition of the
function $\varphi(x,y)$ is the triangular region delimited by the
three lines $y=0$, $y=x$, and $y=1-x$.  Let us consider the ellipse
defined by the equation
\begin{equation}\label{ellipse}
\frac{(1-x-y)^2}{\alpha}+\frac{(x-y)^2}{1-\alpha}=1.
\end{equation}
We denote by $\mathcal{D}_{\mathrm{I}}$ and
$\mathcal{D}_{\mathrm{II}}$ the two regions of the triangle that are
inside, respectively outside, the ellipse, see
Fig.~\ref{fig-ArcticEllipse}. The arc of the ellipse \eqref{ellipse} lying in the
triangle, and separating these two regions, is described by the equation:
\begin{equation}\label{arc}
\sqrt{y}=\sqrt{(1-x)(1-\alpha)}-\sqrt{x\alpha},
\qquad 0\leq y\leq x \leq 1-\alpha.
\end{equation}
We denote this arc by $\mathcal{A}$.

\begin{theorem}\label{Th1}
The function  $\varphi(x,y)$, $(x,y)\in[0,1]\times[0,1]$, is
\begin{equation}\label{sigmazero}
\varphi(x,y)=0,\qquad
(x,y)\in\mathcal{D}_\mathrm{I},
\end{equation}
and
\begin{align}\label{mainresult}
\varphi(x,y)
&=xy\log\frac{h}{\eta}
-\frac{(1-x-y)^2}{2}\log\frac{1-h}{1-\eta}
-\frac{1}{2}\log\frac{1+h}{1+\eta}
\notag\\ &\quad
+(1-x)y\log\frac{y+(1-x)h}{y+(1-x)\eta}
+x(1-y)\log\frac{x+(1-y)h}{x+(1-y)\eta}
\notag\\ &\quad
-(1-x)x\log\frac{x+(1-x)h}{x+(1-x)\eta}
-(1-y)y\log\frac{y+(1-y)h}{y+(1-y)\eta}
\notag\\ &\quad
-\frac{(x-y)^2}{2}\log\frac{x+y+(2-x-y)h}{x+y+(2-x-y)\eta},
\qquad (x,y)\in\mathcal{D}_\mathrm{II},
\end{align}
where
\begin{equation}\label{hxy}
h=h(x,y):=\sqrt\frac{xy}{(1-x)(1-y)}
\end{equation}
and $\eta=\eta(x,y;\alpha)$ is such that
\begin{equation}\label{eta-eq}
\eta\in[0,1],\qquad
\alpha\frac{(1+\eta)^2\big(x+(1-x)\eta\big)\big(y+(1-y)\eta\big)}
{(1-\eta)^2\big(y+(1-x)\eta\big)\big(x+(1-y)\eta\big)}
=1.
\end{equation}
\end{theorem}
The fact that $\varphi(x,y)$ vanishes in the domain
$\mathcal{D}_{\mathrm{I}}$ was already proven in \cite{J-00}.  A few
comments are in order.
\begin{remark}
The function $\varphi(x,y)$ is a continuous function of its
variables. In particular it vanishes on the curve $\mathcal{A}$:
\begin{equation}\label{sigma0}
\varphi(x,y)\big|_{(x,y)\in\mathcal{A}}=0.
\end{equation}
\end{remark}
This property can be verified as follows. Let $(x,y)\in \mathcal{A}$,
then $\alpha=\ac$, where
\begin{equation}\label{avw}
\ac=\ac(x,y):=\big(\sqrt{(1-x)(1-y)}-\sqrt{xy}\big)^2.
\end{equation}
Setting $\alpha=\ac$ in \eqref{eta-eq}, one can directly check
that the root of equation \eqref{eta-eq} is $\eta=h$, with $h$
given by \eqref{hxy}. In other words,
\begin{equation}\label{etah}
\eta(x,y;\alpha)\big|_{(x,y)\in \mathcal{A}}=\eta(x,y;\ac)=h(x,y),
\end{equation}
and \eqref{sigma0} follows from \eqref{mainresult}.

The function $\varphi(x,y)=\varphi(x,y;\alpha)$ may also be viewed as
a function of $\alpha$ at fixed $x$ and $y$.
\begin{remark}
The function $\varphi(x,y)$ satisfies
\begin{equation}
\varphi(x,y;\ac)=0.
\end{equation}
\end{remark}
This property follows from the second equality in \eqref{etah}.
In fact, stronger properties hold.
\begin{proposition}\label{propa3}
Given $(x,y)$, for values of $\alpha$ in the vicinity of
$\ac=\ac(x,y)$, we have
\begin{equation}
\varphi(x,y)=
\begin{cases}
0 & \alpha< \ac
\\
C (\alpha-\ac)^3+O\left((\alpha-\ac)^4\right) & \alpha > \ac,
\end{cases}
\end{equation}
where $C=C(x,y)>0$.
\end{proposition}

\begin{proposition}\label{propt3}
Given a point $(x,y)\in\mathcal{D}_{\mathrm{II}}$, at a
distance $t$ from the arc of ellipse $\mathcal{A}$, one has
\begin{equation}
\varphi(x,y)\propto t^3,\qquad t\to 0^+.
\end{equation}
where the proportionality constant is positive.
\end{proposition}
These properties follows from the explicit form of function $\varphi(x,y)$
given in Theorem~\ref{Th1}.

\subsection{Discussion}

The ellipse \eqref{ellipse} is easily recognized as the Arctic ellipse
of the domino tilings of the original, non-deformed Aztec diamond
\cite{JPS-98}, or, equivalently, of the six-vertex model with domain
wall boundary condition, at its free-fermion point, on the original
square domain.

The function $\varphi(x,y)$ vanishes at $\alpha=\ac$, together with
its first and second derivatives with respect to $\alpha$, and has
non-vanishing (and finite) third-order derivative.  This property
obviously holds also for the free energy $f(x,y)$, see \eqref{fdef}.
Recalling that the parameter $\alpha$ is directly related to the
Boltzmann weights, we thus observe the occurrence of a third-order
phase transition at $\alpha=\ac$.

An alternative point of view is that of considering $\alpha$ fixed,
while varying the shape of the L-shaped domain, and thus the
coordinates $x$ and $y$.  The function $\varphi(x,y)$, studied as a
function of the parameter $t$, again undergoes a third-order phase
transition.

This result leads to an alternative interpretation of the Arctic
ellipse, as a critical curve in the space of the parameters of the
model, a point of view already advocated in \cite{RP-06}. The result
extends that reported in \cite{CP-13}, where the analysis was
restricted to the line $x=y$, and only the critical point
$(x,y)=\big(\frac{1-\sqrt\alpha}{2},\frac{1-\sqrt\alpha}{2}\big)$
could be observed.

The vanishing of the function $\varphi(x,y)$ in domain
$\mathcal{D}_{\mathrm{I}}$ is somewhat obvious, since it is known that
in $\mathcal{D}_{\mathrm{I}}$, the emptiness formation probability is
equal to $1$ up to exponentially small corrections, which are related
to the so-called upper tail rate function \cite{J-00}. The function
$\varphi(x,y)$ in domain $\mathcal{D}_{\mathrm{II}}$ defines instead
the so-called lower tail rate function, whose explicit form is given
by \eqref{mainresult}.

In the derivation of Theorem \eqref{Th1} we rely heavily on
representation \eqref{Irsq}, that is a Coulomb gas on a discrete
lattice, confined by a non-polynomial potential with two hard walls.
Thus, a by-product of our results is that
the hard wall in the discrete Coulomb gas
can induce a soft-edge/hard-edge transition with a
third-order discontinuity in the corresponding free energy.  This is a
well-known phenomenon in the case of a continuous Coulomb gas, see
\cite{MS-14} and references therein. However, in the discrete case,
phase transitions are rather of the Douglas-Kazakov type \cite{DK-93},
while the soft-edge/hard-edge transition is usually
\emph{transparent}, in that it does not come with any discontinuity in
the free energy (see \cite{BKMM-07} for numerous examples).

The peculiar behaviour observed here can be ascribed to the form of
the potential in the vicinity of the hard walls.  In previously
studied cases, in the vicinity of a hard wall situated, say, at $x_0$,
the derivative of the potential has a divergent behaviour $V'(x)\simeq
\log|x-x_0|$, implying a smooth matching of the potential with the
hard wall. In the present model, instead, such smooth matching occurs
only in correspondence of the left hard wall.
As a result, among  the two
possible soft-edge/hard-edge transitions in the present model only
one is transparent, that is related to the left hard wall.
The other one, related to the right hard wall, comes indeed with a third-order phase
transition in the corresponding free energy.

\subsection*{ Acknowledgments}

We thank A. Kuijlaars and P. Zinn-Justin for useful
discussions.  This work is partially supported by the IRSES grant of
EC-FP7 Marie Curie Action ``Quantum Integrability, Conformal Field
Theory and Topological Quantum Computation''.  A.G.P.  acknowledges
partial support from the Russian Foundation for Basic Research, grant
13-01-00336, and from INFN, Sezione di Firenze.

\section{The equilibrium measure}

We first outline some general aspects of the asymptotic behaviour
of the discrete Coulomb gas, along the lines developed in
\cite{BKMM-07}, see also \cite{J-00,BL-13}, and next switch to the
discussion of explicit solutions for the resolvent of the equilibrium
measure.  The end-point equations and the free energy are discussed in
the next section.

\subsection{The discrete Coulomb gas}

In order to evaluate the limit \eqref{sigmadef} and prove
Theorem~\ref{Th1}, we study the asymptotic behaviour of the discrete
Coulomb gas \eqref{Irsq} in the scaling limit, that is
$r,s,q\to\infty$, while preserving the aspect ratio of the L-shaped
domain.  Let $r=[Rs]$, $R\geq 1$ fixed and $q=[Qs]$, $Q\geq 0$ fixed.
We set
\begin{equation}\label{defPhi} \Phi(R,Q):=\lim_{s\to\infty}\frac{1}{s^2}
\log I_{r,s,q}.
\end{equation}
It was shown in \cite{J-00}, see Theorem 2.2 therein, that the above
limit exists.  The function $\Phi(R,Q)$ is related to the function
$\varphi(x,y)$ introduced in \eqref{sigmadef} by
\begin{equation}\label{sigmaPhi}
(1-x)^2\varphi(x,y)=-xy\log(1-\alpha)+y^2\log \sqrt{\alpha}-y^2
\Phi\left(\frac{1-x}{y}, \frac{x-y}{y}\right),
\end{equation}
see \eqref{Irsqdef}. To obtain relation \eqref{sigmaPhi} one
has to take into account the relations
\begin{equation}\label{RQdef}
R=\frac{1-x}{y},\qquad
Q=\frac{x-y}{y},
\end{equation}
which hold in the limit $s\to\infty$.

In \cite{J-00}, the focus was on the distribution of the position of
the rightmost charge, $\mathrm{max}_{1\leq j \leq s}m_j$, that was
shown to follow, in the limit $s\to\infty$, the Tracy-Widom
distribution \cite{TW-94a}. In \cite{J-00}, the explicit form of the
upper tail rate function was evaluated. Our aim here is to evaluate
the explicit form of function $\Phi(R,Q)$, and hence, the lower tail
rate function $\varphi(x,y)$.

\begin{figure}
\centering

\psset{unit=8pt}
\begin{pspicture}(-2,-1)(23,11)
\psset{xunit=2,yunit=2,linewidth=.1}
\savedata{\uno}[
{{0,0}, {0.05, -0.166359}, {0.1, -0.265785}, {0.15, -0.341322}, {0.2,
-0.402044}, {0.25, -0.452216}, {0.3, -0.494321}, {0.35, -0.529977},
{0.4, -0.560319}, {0.45, -0.586179}, {0.5, -0.608198}, {0.55,
-0.626875}, {0.6, -0.642613}, {0.65, -0.655742}, {0.7, -0.666537},
{0.75, -0.675229}, {0.8, -0.682013}, {0.85, -0.687059}, {0.9,
-0.690514}, {0.95, -0.692506}, {1., -0.693147}, {1.05, -0.692537},
{1.1, -0.690765}, {1.15, -0.68791}, {1.2, -0.684044}, {1.25,
-0.67923}, {1.3, -0.673526}, {1.35, -0.666986}, {1.4, -0.659658},
{1.45, -0.651585}, {1.5, -0.642808}, {1.55, -0.633365}, {1.6, -0.623288},
{1.65, -0.612611}, {1.7, -0.601362}, {1.75, -0.589567},
{1.8, -0.577253}, {1.85, -0.564443}, {1.9, -0.551159}, {1.95,
-0.537421}, {2., -0.523248}, {2.05, -0.508659}, {2.1, -0.493669},
{2.15, -0.478295}, {2.2, -0.462553}, {2.25, -0.446455}, {2.3,
-0.430015}, {2.35, -0.413245}, {2.4, -0.396158}, {2.45, -0.378765},
{2.5, -0.361076}, {2.55, -0.343101}, {2.6, -0.324849}, {2.65,
-0.306331}, {2.7, -0.287554}, {2.75, -0.268527}, {2.8, -0.249258},
{2.85, -0.229753}, {2.9, -0.210021}, {2.95, -0.190067}, {3.,
-0.169899}, {3.05, -0.149523}, {3.1, -0.128944}, {3.15, -0.108168},
{3.2, -0.0872014}, {3.25, -0.0660486}, {3.3, -0.0447148}, {3.35,
-0.0232047}, {3.4, -0.0015231}, {3.45, 0.0203256}, {3.5,
  0.0423372}, {3.55, 0.0645076}, {3.6, 0.0868327}, {3.65,
  0.109309}, {3.7, 0.131932}, {3.75, 0.154699}, {3.8,
  0.177607}, {3.85, 0.200652}, {3.9, 0.22383}, {3.95, 0.247139}, {4.,
  0.270577}, {4.05, 0.294139}, {4.1, 0.317823}, {4.15,
  0.341627}, {4.2, 0.365548}, {4.25, 0.389584}, {4.3,
  0.413731}, {4.35, 0.437989}, {4.4, 0.462353}, {4.45,
  0.486823}, {4.5, 0.511396}, {4.55, 0.53607}, {4.6, 0.560843}, {4.65,
   0.585713}, {4.7, 0.610678}, {4.75, 0.635737}, {4.8,
  0.660887}, {4.85, 0.686127}, {4.9, 0.711455}, {4.95, 0.736869}, {5.,
   0.762369}, {5.05, 0.787951}, {5.1, 0.813616}, {5.15,
  0.839361}, {5.2, 0.865185}, {5.25, 0.891086}, {5.3,
  0.917064}, {5.35, 0.943116}, {5.4, 0.969242}, {5.45, 0.99544}, {5.5,
   1.02171}, {5.55, 1.04805}, {5.6, 1.07446}, {5.65, 1.10093}, {5.7,
  1.12748}, {5.75, 1.15408}, {5.8, 1.18076}, {5.85, 1.20749}, {5.9,
  1.23429}, {5.95, 1.26115}, {6., 1.28807}, {6.05, 1.31505}, {6.1,
  1.34209}, {6.15, 1.36918}, {6.2, 1.39633}, {6.25, 1.42354}, {6.3,
  1.45081}, {6.35, 1.47813}, {6.4, 1.5055}, {6.45, 1.53292}, {6.5,
  1.5604}, {6.55, 1.58793}, {6.6, 1.6155}, {6.65, 1.64313}, {6.7,
  1.67081}, {6.75, 1.69854}, {6.8, 1.72631}, {6.85, 1.75413}, {6.9,
  1.782}, {6.95, 1.80991}, {7., 1.83787}, {7.05, 1.86587}, {7.1,
  1.89392}, {7.15, 1.92201}, {7.2, 1.95014}, {7.25, 1.97832}, {7.3,
  2.00654}, {7.35, 2.0348}, {7.4, 2.06309}, {7.45, 2.09143}, {7.5,
  2.11981}, {7.55, 2.14823}, {7.6, 2.17669}, {7.65, 2.20519}, {7.7,
  2.23372}, {7.75, 2.26229}, {7.8, 2.2909}, {7.85, 2.31954}, {7.9,
  2.34822}, {7.95, 2.37694}, {8., 2.40569}, {8.05, 2.43447}, {8.1,
  2.46329}, {8.15, 2.49215}, {8.2, 2.52103}, {8.25, 2.54996}, {8.3,
  2.57891}, {8.35, 2.60789}, {8.4, 2.63691}, {8.45, 2.66596}, {8.5,
  2.69504}, {8.55, 2.72415}, {8.6, 2.75329}, {8.65, 2.78247}, {8.7,
  2.81167}, {8.75, 2.8409}, {8.8, 2.87016}, {8.85, 2.89945}, {8.9,
  2.92877}, {8.95, 2.95812}, {9., 2.98749}, {9.05, 3.0169}, {9.1,
  3.04633}, {9.15, 3.07579}, {9.2, 3.10527}, {9.25, 3.13478}, {9.3,
  3.16432}, {9.35, 3.19388}, {9.4, 3.22347}, {9.45, 3.25309}, {9.5,
  3.28273}, {9.55, 3.3124}, {9.6, 3.34209}, {9.65, 3.3718}, {9.7,
  3.40154}, {9.75, 3.4313}, {9.8, 3.46109}, {9.85, 3.4909}, {9.9,
  3.52074}, {9.95, 3.55059}, {10., 3.58047}}
]

\rput(0,0){\dataplot{\uno}}

\psset{unit=1,linewidth=.075}
\psline{<->}(0,11)(0,0)(23,0)
\psline[linestyle=dashed](20,0)(20,7)
\psline[linewidth=.1](20,7)(20,10)
\rput[rB](-.5,-1.5){$0$}
\rput[B](20,-1.5){$R$}
\rput[B](23,-1.5){$\mu$}
\rput[r](-.5,10){$V(\mu)$}

\end{pspicture}
\caption{The potential $V(\mu)$
with two hard walls, at $\mu=0$ and $\mu=R$.}
\label{fig-Vmu}
\end{figure}
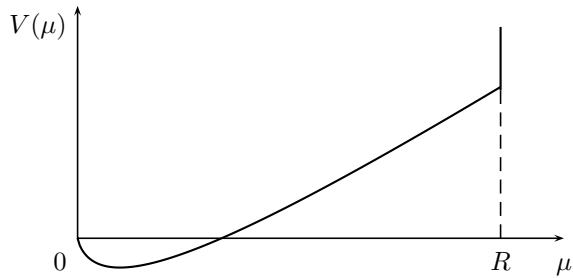

As discussed in \cite{J-00} to evaluate the continuous limit of the
discrete Coulomb gas \eqref{Irsq} we may just as well consider the
corresponding continuous Coulomb gas, in a continuous potential
derived from the original discrete measure. Set
\begin{equation}
V_s^{}(\mu):=-\frac{1}{s} \log \left[\binom{q+s\mu}{q}\alpha^{s \mu}\right],
\qquad \mu\geq 0.
\end{equation}
Using Stirling formula, we have
\begin{equation}\label{potential}
V(\mu):=\lim_{s\to\infty}  V_s(\mu)=-\mu\log\alpha
+\mu\log\mu-(Q+\mu)\log(Q+\mu)+Q\log Q,
\end{equation}
uniformly on compact subsets of $[0,\infty)$.  Note that the
confining potential $V(\mu)$ is complemented by two hard walls, at
$\mu=0$ and $\mu=R$, see Fig.~\ref{fig-Vmu}. We emphasize the
difference in the behaviour of the potential in the vicinity of the
two hard walls: the derivative of the potential diverges
logarithmically at $\mu=0$, while it stays \emph{finite} at $\mu=R$. As a
result, the potential joins up the hard wall smoothly on the left, but
with a corner on the right.

To proceed, we rely on some known results from weighted potential
theory in the case of a discrete Coulomb gas.  This has been discussed
both in the physical \cite{DK-93} and mathematical literature
\cite{DS-97,Kui-00}; see \cite{BKMM-07} for a general mathematical
treatment.  In a nutshell, under mild assumptions on the potential
that are clearly satisfied in our problem, see \cite{J-00}, the
asymptotic behaviour of the discrete Coulomb gas is governed by a
constrained equilibrium problem analogous to that for the continuous
Coulomb gas, with the additional constraint that the particle density
is always less than or equal to $1$, as a consequence of discreteness.

More specifically, the problem reduces to the evaluation of the
equilibrium measure, $\rho_0(\mu)\rmd\mu$, that is the unique minimizer
of the functional
\begin{equation}\label{functional}
S[\rho]=-\int\int_{\mu\not =\nu}\log\vert\mu-\nu\vert\rho(\mu)
\rho(\nu)\,\rmd\mu\rmd\nu+\int V(\mu)\rho(\mu)\,\rmd\mu,
\end{equation}
subject to the constraint (normalization condition)
\begin{equation}\label{normalize}
\int\rho(\mu)\,\rmd\mu=1,
\end{equation}
and to the additional constraint
\begin{equation}\label{leqone}
\rho(\mu)\leq 1,
\end{equation}
that is imposed by the discreteness of the measure of the Coulomb gas.
In the above formulas, integration is understood over the interval
$[0,R]$.

It is known that, due to the presence of the last constraint, the
equilibrium measure partitions the interval $[0,R]$ into three sets,
$I_-$, $I_0$, $I_+$, according to $\rho_0(\mu)=0$, $0<\rho_0(\mu)<1$,
or $\rho_0(\mu=1$, respectively. Each set is the union of a finite
number of intervals, named \emph{voids}, \emph{bands}, or
\emph{saturated regions}, respectively. Voids and saturated regions
are collectively named \emph{gaps}.

\subsection{The resolvent}

The most convenient way to evaluate the equilibrium measure is by
introducing its \emph{resolvent}
\begin{equation}\label{Wdef}
W(z)=\int_{0}^R \frac{\rho_0(\mu)}{z-\mu}\, \rmd \mu,
\qquad z\notin [0,R],
\end{equation}
that has the following properties:
\begin{enumerate}
\item $W(z)$ is analytic in $\mathbb{C}\setminus [0,R]$.
\item For large $z$, the resolvent has the asymptotic behaviour
\begin{equation}\label{largez}
W(z)= \frac{1}{z}+\frac{E}{z^2}+O\left(z^{-3}\right),
\qquad \vert z\vert\to\infty,
\end{equation}
where the leading coefficient is determined by the normalization
condition \eqref{normalize}, and $E$ is
the first moment of the equilibrium measure:
\begin{equation}\label{firstmoment}
E=\int_0^R\mu\, \rho_0(\mu)\,\rmd\mu.
\end{equation}
\item The particle density can be found as the discontinuity of the
resolvent across its cuts, $I_0\cup I_+$,
\begin{equation}\label{rhodef}
\rho_0(\mu)=-\frac{1}{2\pi\rmi}\left[W(\mu+\rmi 0)-W(\mu-\rmi 0)\right],
\qquad \mu\in [0,R].
\end{equation}
In particular, it follows that
\begin{equation}
W(\mu+\rmi 0)-W(\mu-\rmi 0)
=
\begin{cases}
0& \mu\in I_-
\\
-2\pi \rmi &  \mu\in I_+.
\end{cases}
\end{equation}
\item The resolvent satisfies
\begin{equation}\label{WWV}
W(\mu+\rmi 0)+W(\mu+\rmi 0)= V'(\mu), \qquad \mu\in I_0.
\end{equation}
\end{enumerate}
The above properties, together with some `ansatz' on the structure of
the partition of the interval $[0,R]$ into voids, bands, and
saturation regions, are sufficient to completely determine the
resolvent associated to the considered discrete Coulomb gas.

Before discussing the issues related to the discrete nature of the
Coulomb gas, and to the related phenomenon of emergence of saturated
regions, it is useful to recall the procedure for the determination of
the resolvent in a case where the discreteness and hard walls play no
role, such as that of a single band $[a,b]$, with two voids, $[0,a]$
and $[b,R]$.  The equation for the resolvent is a singular integral
equation of the form \eqref{WWV}, where $I_0 = [a,b]$.  The end-points
of the band are determined by the condition that the solution of
\eqref{WWV} should match at leading order the large $z$ behaviour
$W(z)\sim 1/z$, dictated by \eqref{largez}.

However, if the density found by solving \eqref{WWV} does not satisfy
the condition \eqref{leqone}, or the end-points $a$ and $b$ are not
internal to the interval $[0,R]$, then one has to introduce some
suitable ansatz about the structure of the partition of the interval
$[0,R]$ into voids, bands, and saturated regions, with $I_+$
consisting of at least one interval.

The existence of at least one saturated region (that is, $I_+\not
=\emptyset$) implies the following form of the resolvent:
\begin{equation}
W(z)=\int_{I_+}^{}\frac{1}{z-\mu}\,\rmd\mu+ H(z),
\qquad
H(z):=\int_{I_0}^{}\frac{\rho_0(\mu)}{z-\mu}\,\rmd\mu.
\end{equation}
Here the function $H(z)$ can be found from the equation
\begin{equation}\label{spegen}
H(\mu+\rmi 0)+H(\mu-\rmi 0)=U(\mu), \qquad    \mu\in I_0,
\end{equation}
where
\begin{equation}
U(z)=-2\int_{I_+}^{}\frac{1}{z-\mu}\,\rmd\mu+V'(z).
\end{equation}
Assuming that $I_0$ consists of just a single interval (the so-called
one-band ansatz, \cite{BKMM-07}), with the end-points $a$ and $b$
satisfying $0< a< b<R$, the solution of \eqref{spegen} such that
$H(z)=O(1/z)$ as $z\to\infty$, is
\begin{equation}\label{Wint}
H(z)=\frac{\sqrt{(z-a)(z-b)}}{2\pi}
\int_a^b\frac{U(\mu)}{(z-\mu)\sqrt{(\mu-a)(b-\mu)}}\,\rmd \mu.
\end{equation}
The end-points $a$ and $b$ can be found by solving the equations on
the coefficients of order $z^0$ and $z^{-1}$ terms of the large $z$
expansion of \eqref{Wint}, to match the the large $z$ behaviour
\eqref{largez} at leading order. Finally, the explicit form of the
particle density is derived by using \eqref{rhodef}, and its
consistency against condition \eqref{leqone} may be verified.

In formulating some ansatz on the structure of the partition of the
interval $[0,R]$ into voids, bands, and saturation regions, we
take into account the form of the potential \eqref{potential},
together with the presence of two hard walls (that is infinite
potential barriers) at $\mu=0,R$. It is clear that the band-gap
structure may be affected by varying the parameters $R\in[1,\infty)$
and $Q\in[0,\infty)$.  The whole solution is obtained by investigating
how the band-gap structure is modified as these parameters are varied.

Concerning the shape of the potential \eqref{potential}, it is clear
that $V(0)=0$ and, as $\mu\to\infty$, the function $V(\mu)$ is
asymptotically linear, $V(\mu)\sim-(\log\alpha) \mu$, with a positive
slope (recall that $0<\alpha<1$).  The derivative of the potential
reads
\begin{equation}\label{Vprime}
V'(\mu)=-\log\alpha+\log\mu-\log(\mu+Q).
\end{equation}
Thus the function $V(\mu)$ has a single minimum in the interval
$[0,R]$, at $\mu=\mu_0$, where
\begin{equation}\label{mu0}
\mu_0=\frac{Q \alpha}{1-\alpha}.
\end{equation}
A nonvanishing density (with bands and possibly saturated regions) is
thus expected in the vicinity of $\mu_0$.

Concerning the presence of hard walls, it is known that these affect
the band-gap structure in that they admit only gaps in their vicinity
\cite{BKMM-07}.  Thus, in our case, the intervals adjacent to the points $\mu=0$ and
$\mu=R$ can only be voids or saturated regions.  It is clear, basing
on the physical picture of the Coulomb gas, that the presence of a
void or a saturated region is related to the distance of the minimum
$\mu_0$ from the hard walls at $\mu=0,R$.

A detailed analysis provided below shows that four possible band-gap
structures, or \emph{scenarios} may occur.  We refer to them as
Regime IA, Regime IB, Regime IIA, and Regime IIB for definiteness.  In
all these scenarios, there is just one band, that we denote $[a,b]$,
between two gaps, $[0,a]$ and $[b,R]$. The four scenarios differ one
another in the nature (void or saturated region) of the two gaps.
Namely, in Regime IA the first gap, $[0,a]$, is a saturated region,
while the second gap, $[b,R]$, is a void; in Regime IB both gaps are
voids; in Regime IIA both gaps are saturated regions; finally, in
Regime IIB the first gap is a saturated region while the second is a
void. The four regimes correspond to four different
regions in the domain $Q\geq 0$, $R\geq 1$.

\subsection{Regime I}

Regime IA and Regime IB are closely related. Actually, they can be
viewed as two different scenarios for one same regime, Regime I,
characterized by a void on the rigth gap $[b,R]$.  The equilibrium
measures for Regime IA and Regime IB were first derived in \cite{J-00}.

\begin{proposition}\label{regime1a}
The resolvent of the equilibrium measure for the continuum limit of
the discrete Coulomb gas \eqref{Irsq} in Regime IA is
\begin{multline}\label{WIA}
W(z)=-\log\sqrt{\alpha}
-\log\frac{\sqrt{a(z-b)}+\sqrt{b(z-a)}}{\sqrt{(b-a)z}}
\\
-\log\frac{\sqrt{(a+Q)(z-b)}+
\sqrt{(b+Q)(z-a)}}{\sqrt{(b-a)z}}.
\end{multline}
where the parameters $a$ and $b$, the two band end-points, are
\begin{equation}\label{bq1}
a=\frac{\big(1-\sqrt{\alpha(1+Q)}\big)^2}{1-\alpha},\qquad
b=\frac{\big(1+\sqrt{\alpha(1+Q)}\big)^2}{1-\alpha}.
\end{equation}
\end{proposition}

In Regime IA, there is one band $[a,b]$, with a saturated region
$[0,a]$, and a void $[b,R]$.  The saturated region in the interval
$[0,a]$ implies that the resolvent has the form
\begin{equation}
  W(z)=\log \frac{z}{z-a}+H(z),\qquad
  H(z):=\int_{a}^{b}\frac{\rho_0(\mu)}{z-\mu}\, \rmd \mu.
\end{equation}
The function $W_\text{IA}(z)$ should solve equation \eqref{spegen}
with $U(z)=V'(z)$ and $I_0=[a,b]$, where the function $H(z)$ should
solve \eqref{spegen}, with function $U(\mu)$ chosen as
\begin{equation}
U(\mu)=-\log\alpha+\log\frac{(\mu-a)^2}{\mu(\mu+Q)}.
\end{equation}
Evaluating the integral in \eqref{Wint} with this function and coming
back to function $W(z)$, we obtain expression \eqref{WIA}.

The density in the interval $[a,b]$ is given by the expression
\begin{equation}
\rho_0(\mu)=
\frac{1}{\pi}\arctan\sqrt{\frac{a(b-\mu)}{b(\mu-a)}} +
\frac{1}{\pi}\arctan\sqrt{\frac{(a+Q)(b-\mu)}{(b+Q)(\mu-a)}},
\end{equation}
while, by construction, $\rho_0(\mu)=1$, for $\mu\in [0,a]$, and
$\rho_0(\mu)=0$, for $\mu\in [b,R]$.

Requiring $W(z)\sim 1/z$, as $z\to\infty$, from \eqref{WIA} we
obtain the following equations for the end-points $a$ and $b$:
\begin{equation}\label{eqRIA}
\frac{\sqrt{b+Q}-\sqrt{a+Q}}{\sqrt{b}+\sqrt{a}}=\sqrt{\alpha},\qquad
\frac{\sqrt{(b+Q)(a+Q)}+\sqrt{a b}-Q}{2}=1.
\end{equation}
These equations can be easily solved, with the result \eqref{bq1}

It is clear from qualitative arguments that the current regime holds
for relatively small values of $Q$, and relatively large values of
$R$, so that the minimum of the potential is relatively close to the
left hard wall at origin, and relatively far from the right hard wall
at $R$.

Imposing $a=0$, we get the value $Q=\Qc$ corresponding to the
soft-edge/hard edge transition induced by the left hard wall at
origin. We have
\begin{equation}\label{Qc}
\Qc=\frac{1-\alpha}{\alpha}.
\end{equation}
It is clear that Regime IA corresponds to values of $Q<\Qc$.

Imposing $b=R$, we get the value $R=\Rc$ corresponding to the
soft-edge/hard edge transition induced by the right hard wall located
at $R$. We have
\begin{equation}\label{Rc}
\Rc=\frac{\big(1+\sqrt{\alpha(1+Q)}\big)^2}{1-\alpha}.
\end{equation}
Clearly, this is the minimum value of $R$ for which the scenario of
Regime IA is applicable.

\begin{proposition}\label{regime1b}
The resolvent of the equilibrium measure for the continuum limit of
the discrete Coulomb gas \eqref{Irsq} in Regime IB is
\begin{equation}\label{WIB}
W(z)=-\log\sqrt{\alpha}
+\log\frac{\sqrt{a(z-b)}+\sqrt{b(z-a)}}{\sqrt{(a+Q)(z-b)}+
\sqrt{(b+Q)(z-a)}}.
\end{equation}
where the parameters $a$ and $b$, the two band end-points, are given again by
\eqref{bq1}
\end{proposition}

In Regime IB, there is one band $[a,b]$ between two voids, $[0,a]$ and
$[b,R]$. The resolvent $W(z)$ is simply given by the solution of
\eqref{spegen} with $U(z)=V'(z)$ and $I_0=[a,b]$. As a result, after
evaluating the integral \eqref{Wint}, we get expression \eqref{WIB}.

The density in the band is given by
\begin{equation}
\rho_0(\mu)=
-\frac{1}{\pi}\arctan\sqrt{\frac{a(b-\mu)}{b(\mu-a)}} +
\frac{1}{\pi}\arctan\sqrt{\frac{(a+Q)(b-\mu)}{(b+Q)(\mu-a)}},\qquad
\mu \in[a,b],
\end{equation}
while, by construction, $\rho_0(\mu)=0$, for $\mu\in [0,a]\cup [b,R]$.

Requiring $W(z)\sim1/z$, as $z\to\infty$, from
\eqref{WIB} we have the following
equations for the end-points $a$ and $b$:
\begin{equation}
\frac{\sqrt{b+Q}-\sqrt{a+Q}}{\sqrt{b}-\sqrt{a}}
=\sqrt{\alpha},
\qquad
\frac{\sqrt{(a+Q)(b+Q)}-\sqrt{ab}-Q}{2}=1.
\end{equation}
Note that these equations differ from those in Regime IA, see
\eqref{eqRIA}, just by the sign of $\sqrt{a}$. Their formal
solution, satisfying $0<a<b$, is given again by
\eqref{bq1}, in which now $Q>\Qc$, see \eqref{Qc}.

We also note that the critical value of $R$, namely, its minimal
value, for which the Regime IB is applicable, is given by \eqref{Rc}.
Thus the minimum value of $R$ for which the
scenario of Regime IB is applicable is the same as for Regime IA.

In Regime IB, all charges accumulate in the vicinity of the minimum
$\mu_0$, and one should check that the obtained solution for
\eqref{spegen} with $U(z)=V'(z)$ and $I_0=[a,b]$ satisfies condition
\eqref{leqone}.  This is indeed the case, thus ensuring that no
saturated region arises.

Summarizing, a critical value $\Qc$ exists, that separates the two
regimes, with Regime IA corresponding to $Q<\Qc$ and Regime IB to
$Q>\Qc$.  The value $\Qc$ is determined by the condition $a=0$, that
is by the vanishing of the left gap, and the corresponding
soft-edge/hard-edge transition. Similarly, a critical value $\Rc$,
determined by the condition $R=b$, exists, such that Regime IA and IB
correspond to $R>\Rc$. The value $\Rc$ is determined by the condition
$b=R$, that is by the vanishing of the right gap, and the
corresponding soft-edge/hard-edge transition.

\subsection{Regime II}

Regime IIA and Regime IIB are closely related. Actually, they can be
viewed as two different scenarios for one same regime, Regime II,
characterized by a saturated region on the rigth gap $[b,R]$. They
both correspond to values $R<\Rc$, with $\Rc$ given by \eqref{Rc}.

\begin{proposition}\label{regime2a}
 The resolvent of the equilibrium measure for the continuum limit of
  the discrete Coulomb gas \eqref{Irsq} in Regime IIA is
\begin{multline}\label{WIIA}
W(z)=-\log\sqrt{\alpha}+\log\frac{z}{z-R}
-\log\frac{\sqrt{a(z-b)}+\sqrt{b(z-a)}}
{\sqrt{(R-a)(z-b)}+\sqrt{(R-b)(z-a)}}
\\
-\log\frac{\sqrt{(a+Q)(z-b)}+\sqrt{(b+Q)(z-a)}}
{\sqrt{(R-a)(z-b)}+\sqrt{(R-b)(z-a)}}.
\end{multline}
where the parameters $a$ and $b$, the two band end-points, are the solutions
of the two equations:
\begin{equation}\label{eqRIIA}
\begin{split}
\sqrt{\alpha}
\frac{\sqrt{R-b}-\sqrt{R-a}}{\sqrt{R-b}+\sqrt{R-a}}
\frac{\sqrt{b}+\sqrt{a}}
{\sqrt{b+Q}-\sqrt{a+Q}}
&=1,
\\
\frac{\sqrt{ab}+\sqrt{(a+Q)(b+Q)}-Q}{2}+\sqrt{(R-a)(R-b)}
&=1.
\end{split}
\end{equation}
\end{proposition}

In Regime IIA there is a band in $[a,b]$ between two saturated regions
in $[0,a]$ and $[b,R]$.  This implies that the resolvent in this case
has the form
\begin{equation}
W(z)=\log \frac{z}{z-a}+\log\frac{z-b}{z-R}+H(z),\qquad
H(z):=\int_{a}^{b}\frac{\rho_0(\mu)}{z-\mu}\, \rmd \mu.
\end{equation}
The function $H(z)$ should solve equation \eqref{spegen},
where the function $U(\mu)$ is given by the expression
\begin{equation}
U(\mu)=-\log\alpha+\log\frac{(\mu-a)^2(\mu-b)^2}{\mu(\mu+Q)(\mu-R)^2}.
\end{equation}
Evaluating the integral \eqref{Wint}, and coming back to function
$W(z)$, we get \eqref{WIIA}.

The density is given by
\begin{multline}
\rho_0(\mu)=
1+\frac{1}{\pi}\arctan\sqrt{\frac{a(b-\mu)}{b(\mu-a)}}
+\frac{1}{\pi}\arctan\sqrt{\frac{(a+Q)(b-\mu)}{(b+Q)(\mu-a)}}
\\
-\frac{2}{\pi}\arctan\sqrt{\frac{(R-a)(b-\mu)}{(R-b)(\mu-a)}},\qquad
\mu\in[a,b],
\end{multline}
with  $\rho_0(\mu)=1$, for $\mu\in[0,a]\cup[b,R]$, by construction.

Imposing $W(z)\sim 1/z$, as $z\to\infty$, from \eqref{WIIA} the system
of equations \eqref{eqRIIA} is obtained for the end-points $a$ and
$b$.

\begin{proposition}\label{regime2b}
The resolvent of the equilibrium measure for the continuum limit of
the discrete Coulomb gas \eqref{Irsq} in Regime IIB is
\begin{multline}\label{WIIB}
W(z)=-\log\sqrt{\alpha}
-2\log\frac{\sqrt{(b-a)(z-R)}}{\sqrt{(R-a)(z-b)}
+\sqrt{(R-b)(z-a)}}\\
+\log\frac{\sqrt{a(z-b)}+\sqrt{b(z-a)}}{\sqrt{(a+Q)(z-b)}+
\sqrt{(b+Q)(z-a)}}.
\end{multline}
where the parameters $a$ and $b$, the two band end-points, are the solutions
of the two equations:
\begin{equation}\label{eqRIIB}
\begin{split}
\sqrt{\alpha}
\frac{\sqrt{R-b}-\sqrt{R-a}}{\sqrt{R-b}+\sqrt{R-a}}
\frac{\sqrt{b}-\sqrt{a}}
{\sqrt{b+Q}-\sqrt{a+Q}}
&=1,
\\
\frac{-\sqrt{ab}+\sqrt{(a+Q)(b+Q)}-Q}{2}+\sqrt{(R-a)(R-b)}
&=1.
\end{split}
\end{equation}
\end{proposition}

In Regime IIB there is a band in $[a,b]$, with a saturated region in
$[b,R]$. Therefore the resolvent has the form
\begin{equation}
W(z)=\log\frac{z-b}{z-R}+H(z),\qquad
H(z):=\int_{a}^{b}\frac{\rho_0(\mu)}{z-\mu}\, \rmd \mu.
\end{equation}
The function $H(z)$ should solve equation \eqref{spegen},
where the function $U(\mu)$ reads
\begin{equation}
U(\mu)=-\log\alpha+\log\frac{\mu(\mu-b)^2}{(\mu+Q)(\mu-R)^2}.
\end{equation}
As a result, expression \eqref{WIIB} is obtained for the resolvent.

The density is given by
\begin{multline}
\rho_0(\mu)=
\frac{1}{\pi}\arctan\sqrt{\frac{a(b-\mu)}{b(\mu-a)}}
-\frac{1}{\pi}\arctan\sqrt{\frac{(a+Q)(b-\mu)}{(b+Q)(\mu-a)}}
\\
-\frac{2}{\pi}\arctan\sqrt{\frac{(R-a)(b-\mu)}{(R-b)(\mu-a)}},\qquad
\mu\in[a,b],
\end{multline}
with $\rho_0(\mu)=0$, for $\mu\in[0,a]$, and $\rho_0(\mu)=1$,
for $\mu\in[b,R]$, by  construction.

The condition that the resolvent \eqref{WIIB} has the behaviour
$W(z)\sim 1/z$,  as $z\to\infty$, leads to the system of equations
\eqref{eqRIIB} for the end-points $a$ and $b$.

Just as for the case of Regime I, Regime IIA and Regime IIB
corresponds to values of $Q$ in the ranges $Q<\Qc$ or $Q>\Qc$,
respectively.  The value $\Qc$ is given by the condition that $a=0$,
that is by the vanishing of the left gap, and the corresponding
soft-edge/hard-edge transition.  In Regime II the critical value $\Qc$
has not the simple form \eqref{Qc}, but it is given instead by some
more complicate expression involving the parameter $R$ in a
non-trivial way.

Finally, note that both in Regime IIA and Regime IIB, as the parameter
$R$ is decreased the band $[a,b]$ shrinks down, vanishing in the limit
$R\to 1$.  More precisely, in Regime IIA, one has $a\to b\ne 0$, and
in Regime IIB $a, b\to 0$ as $R\to 1$. In both cases, the density
tends to the saturated value 1 everywhere in the interval $[0,1]$.

\subsection{The Arctic ellipse}

Regime I and Regime II discussed above correspond to values of $R$ in the
interval $R>\Rc$, and $1\leq R <\Rc$, respectively, where $\Rc=\Rc(Q)$
is given by \eqref{Rc}, and $Q$ is allowed to take arbitrary values in
the interval $[0,\infty)$. We discuss here how these conditions
translate in terms of the coordinates $x,y\in[0,1]$.

Clearly, the transition between the two regimes corresponds to
$R=\Rc(Q)$, that gives a relation between the two parameters $R$ and
$Q$, namely
\begin{equation}
\sqrt{(1-\alpha)R}=1+\sqrt{\alpha(1+Q)}.
\end{equation}
Recalling \eqref{RQdef}, this relation translates into the following relation between
the coordinates $x$ and $y$:
\begin{equation}\label{sqrtxy}
\sqrt y =\sqrt{(1-\alpha)(1-x)}-\sqrt{\alpha x}.
\end{equation}
This is readily recognized as the portion of the Arctic ellipes
\eqref{ellipse}, between the two contact points $(x,y)=(1-\alpha,0)$
and $(x,y)=(0,1-\alpha)$.  The additional condition $x\geq y$
(corresponding to the condition $R\geq 0$) selects the arc of ellipse
$\mathcal{A}$, see \eqref{arc}, separating the two domains
$\mathcal{D}_{\mathrm{I}}$ and $\mathcal{D}_{\mathrm{II}}$, see
Fig.~\ref{fig-ArcticEllipse}.

It can also be easily seen that the values of $x$ and $y$, which
corresponds to $R>\Rc$, i.e., to Regime I, are subject to the
conditions $y<\yc$ and $y\leq x$, where $\yc=\yc(x)$ follows from
\eqref{sqrtxy}, and is given by
\begin{equation}
\yc =\begin{cases}
\big(\sqrt{(1-\alpha)(1-x)}-\sqrt{\alpha x}\big)^2
&
x\in[(1-\sqrt\alpha)/2,1-\alpha]
\\
0 & x\in [1-\alpha,1].
\end{cases}
\end{equation}
This conditions implies that in Regime I we have
$(x,y)\in\mathcal{D}_\mathrm{I}$. Similarly, one can verify that
Regime II corresponds to values of the coordinates $x$ and $y$ such
that $(x,y)\in\mathcal{D}_\mathrm{II}$.

Summarizing, Regime I of the Coulomb gas corresponds in the six-vertex
model in an L-shaped domain to situations where the cut-off corner is
sufficiently small to lie totally outside the Arctic ellipse.
Regime II corresponds in the six-vertex model to situations
where the cut-off corner is sufficiently large to overlap with the
Arctic ellipse. As explained in the next section, this transition
manifests as a third-order phase transition in the free energy of the
Coulomb gas, and consequently, of the six-vertex model in an
L-shaped domain. Instead, the transition between Regime IA and Regime IB (or
also between Regime IIA and Regime IIB), although being as well a
soft-edge/hard-edge transition, does not induce any discontinuity in
the Coulomb gas free energy.

\section{The free energy}

To evaluate function $\Phi(R,Q)$, we exploit the specific dependence
of the discrete Coulomb gas \eqref{Irsq} on the variable $\alpha$.
Recall the definition \eqref{firstmoment} of the first moment of the
equilibrium measure. Clearly, we have the relation
\begin{equation}\label{limI}
E=\lim_{s\to\infty}\frac{1}{s^2}
\frac{\partial\log I_{r,s,q}}{\partial\log\alpha}.
\end{equation}
Given the resolvent $W(z)$, the quantity $E$ can be easily extracted
as the $1/z^2$ coefficient of its large $z$ expansion, see
\eqref{largez}.  Using \eqref{defPhi} in \eqref{limI}, we have the
following ordinary first-order differential equation for the function
$\Phi(R)$:
\begin{equation}
\frac{\partial}{\partial\log\alpha} \Phi(R,Q)=E.
\end{equation}
This equation can be easily solved,
\begin{equation}\label{intE}
\Phi(R,Q)=\int E\, \frac{\rmd \alpha}{\alpha}+C(R,Q).
\end{equation}
where $C(R,Q)$ is some function independent of $\alpha$.

To fix the integration constant $C(R,Q)$, we use the fact that
the function $\varphi(x,y)$ admits direct evaluation when the
parameter $\alpha$ takes the limiting values $\alpha=0$ and
$\alpha=1$. These two cases play the role of initial conditions
in determining the function $\varphi(x,y)$ in Regime I and Regime II,
respectively.

\subsection{The limiting cases}

Let us first consider the case $\alpha=0$. To start with, let us come
back to the Coulomb gas at finite values of $r,s,q$.  From
\eqref{Irsq} it follows that the expression $I_{r,s,q}$ is a
polynomial in $\alpha$.  Its lowest order term is of degree
$s(s-1)/2$; the corresponding coefficient can be found (modulo the
number of permutations of $m_j$'s) by setting $m_j=j-1$, that gives
\begin{equation}
I_{r,q,s}\sim \alpha^{s(s-1)/2},
\qquad\alpha\to 0.
\end{equation}
This, together with \eqref{Irsqdef}, implies
\begin{equation}\label{alpha=0a}
F_{r,s,q}\big|_{\alpha=0}=1.
\end{equation}
Hence, by \eqref{sigmadef}, we have, in particular,
\begin{equation}\label{sigmaalpha=0}
\varphi(x,y)\big|_{\alpha=0}=0.
\end{equation}
This condition will allow us to fix the integration constant $C(R,Q)$
in Regime I.

Let us now turn to the case $\alpha=1$. Since the function $F_{r,s,q}$
has a zero at $\alpha=1$ of degree $s(s+q)$, the function
$\varphi(x,y)=\varphi(x,y;\alpha)$ has a logarithmic singularity as
$\alpha\to 1$.  Besides this, the function $\varphi(x,y)$ has a
regular $\alpha\to 1$ part, which is relevant for our analysis.

To compute it, we focus on the Hankel determinant appearing in
$F_{r,s,q}$, see representation \eqref{newEFP}.  In the case
$\alpha=1$, the determinant can be evaluated in closed form.
\begin{proposition}
For arbitrary non-negative integers $r,s,q$, $s\leq r$, the
following holds
\begin{equation}
\det_{1\leq j,k\leq s}
\left[\sum_{m=0}^{r-1}m^{j+k-2}\binom{m+q}{m}\right]
= \prod_{j=0}^{s-1}
\frac{(j!(j+q)!)^2 (j+q+r)!}{q!(r-j-1)!(2j+q)!(2j+q+1)!}.
\end{equation}
\end{proposition}
Indeed, the determinant in the left hand side can be recognized as the
Gram determinant of Hahn polynomials $Q_j(m;q,0,r-1)$. The statement
follows from the standard technique of orthogonal polynomials. For an
account of the properties of Hahn polynomials, see, e.g., the
monograph \cite{KLS-10}, Sect.~9.5.

Therefore, as $\alpha\to1$,
\begin{equation}\label{fata1}
F_{r,s,q}\sim
\prod_{j=0}^{s-1}
\frac{j!(j+q)! (j+q+r)!}{q!(r-j-1)!(2j+q)!(2j+q+1)!}
\cdot (1-\alpha)^{s(s+q)}.
\end{equation}
Let $s+q=[xN]$, $s=[yN]$, with $x,y\in[0,1]$ fixed, $N=r+s+q$. We set
\begin{equation}
\psi(x,y)=-\lim_{N\to\infty}
\frac{1}{N^2} \log\left(\prod_{j=0}^{s-1}
\frac{j!(j+q)! (j+q+r)!}{q!(r-j-1)!(2j+q)!(2j+q+1)!}\right).
\end{equation}
From \eqref{fata1}, recalling \eqref{sigmadef}, it follows that
\begin{equation}\label{limto1}
\lim_{\alpha\to 1} \left[\varphi(x,y)+xy\log(1-\alpha)\right]
=\psi(x,y).
\end{equation}
Using the Stirling formula, we have
\begin{multline}\label{psi}
\psi(x,y)=
\frac{(x+y)^2}{2}\log(x+y)
-\frac{(1-x-y)^2}{2}\log(1-x-y)
\\
-\frac{x^2}{2}\log x
+\frac{(1-x)^2}{2}\log(1-x)
-\frac{y^2}{2}\log y
+\frac{(1-y)^2}{2}\log(1-y).
\end{multline}
Recalling \eqref{sigmaPhi}, relation \eqref{limto1} implies
\begin{equation}
\Phi(R,Q)\big|_{\alpha=1}=-(R+Q+1)^2\psi(x,y).
\end{equation}
Recall that $x$ and $y$ are related to $R$ and $Q$ by
\begin{equation}
x=\frac{1+Q}{R+Q+1},\qquad
y=\frac{1}{R+Q+1},
\end{equation}
see \eqref{RQdef}.  Finally, due to \eqref{psi}, we obtain
the condition
\begin{multline}\label{Phiat1}
\Phi(R,Q)\big|_{\alpha=1}=
-\frac{R^2}{2}\log R+\frac{(R-1)^2}{2}\log (R-1)
-\frac{(R+Q)^2}{2}\log (R+Q)
\\
+\frac{(1+Q)^2}{2}\log(1+Q)
-\frac{(2+Q)^2}{2}\log(2+Q)
\\
+\frac{(R+Q+1)^2}{2}\log(R+Q+1).
\end{multline}
This condition make it possible to fix the integration constant $C(R,Q)$
in Regime II.

After these preliminary remarks, we turn to the evaluation of the free
energy of the discrete Coulomb gas in both regimes.

\subsection{Regime I}

In order to extract the first moment of the equilibrium measure in
Regime IA and Regime IB, we evaluate the term $O(1/z^2)$ in the
$z\to\infty$ expansion of the resolvent, see \eqref{WIA} and
\eqref{WIB}, respectively. In both regimes we obtain
\begin{equation}
E=\frac{2+Q}{8}(a+b)+\frac{Q^2}{4}-\frac{Q}{4}\sqrt{(a+Q)(b+Q)}.
\end{equation}
Substituting here the expressions for the end-points $a$
and $b$, see \eqref{bq1}, we have
\begin{equation}
E=(Q+1)\frac{\alpha}{1-\alpha}+\frac{1}{2}.
\end{equation}
Integrating with respect to $\alpha$, and recalling \eqref{intE}, we get
the free energy of the discrete Coulomb gas for Regime I,
 \begin{equation}\label{PhiRI}
\Phi(R,Q)=-(Q+1)\log(1-\alpha)+\log\sqrt\alpha+C(R,Q),
\end{equation}
where the integration constant $C(R,Q)$ is still to be determined.
The last relation implies
\begin{equation}
\varphi(x,y)=C(R,Q),
\end{equation}
see \eqref{sigmaPhi}.  Finally, the initial condition
\eqref{sigmaalpha=0} determines the value of the integration
constant,
\begin{equation}
C(R,Q)=0. 	
\end{equation}
As a result, we obtain
\begin{equation}
\varphi(x,y)=0,  	
\qquad (x,y)\in\mathcal{D}_\mathrm{I},
\end{equation}
as stated in Theorem \ref{Th1}.

\subsection{Regime II: The end-point equations}

In the previous section we obtained that the end-points $a$ and
$b$ of the band in Regime IIA, see \eqref{eqRIIA}, and Regime IIB, see
\eqref{eqRIIB}, obey the equations
\begin{equation}\label{eqRII}
\begin{split}
\sqrt{\alpha}
\frac{\sqrt{R-b}-\sqrt{R-a}}{\sqrt{R-b}+\sqrt{R-a}}
\frac{\sqrt{b}+\nu\sqrt{a}}
{\sqrt{b+Q}-\sqrt{a+Q}}
&=1,
\\
\frac{\nu\sqrt{ab}+\sqrt{(a+Q)(b+Q)}-Q}{2}+\sqrt{(R-a)(R-b)}
&=1,
\end{split}
\end{equation}
where
\begin{equation}
\nu=
\begin{cases}
+1 &\text{for Regime IIA}
\\
-1 &\text{for Regime IIB}.
\end{cases}
\end{equation}
These are essentially equivalent to quartic equations; a direct
treatment is given in appendix A. The most important concern about the
end-points in the approach we follow here, is that they need to be
represented in such a way that the integral in \eqref{intE} might be
evaluated explicitly.

For this reason we consider here a solution of equations \eqref{eqRII}
in which the second equation is viewed as defining an algebraic curve
in the variables $a$ and $b$, while the first equation is regarded as
fixing a point on this curve. Indeed, by squaring the second equation
twice (since it involves the linear combination of three square roots with
different arguments), $a$ and $b$ are easily seen to lie on an
algebraic curve of degree four.  It turns out that in suitable
variables, this is simply a cubic curve, which is moreover of rational
(rather than elliptic) type. In this construction the first equation
in \eqref{eqRII} becomes a quartic equation for the parameter along
this curve. Furthermore, the solution of this quartic equation can be
viewed as a change of  variables, so that the integration
\eqref{intE} turns out to be feasible.

Instead of dealing with the end-points $a$ and $b$ as unknowns,
we introduce the quantities
\begin{equation}\label{ABC}
\begin{split}
A_\pm
&=\frac{\big(\sqrt{b}\pm\nu\sqrt{a}\big)^2}{4},
\\
B_\pm
&=\frac{\left(\sqrt{b+Q}\pm\sqrt{a+Q}\right)^2}{4},
\\
C_\pm
&=\frac{\left(\sqrt{R-a}\pm\sqrt{R-b}\right)^2}{4}.
\end{split}
\end{equation}
Obviously, the end-points can be expressed in terms of these new
quantities, for example
\begin{equation}
a=A_{+}+A_{-}-2\sqrt{A_{+}A_{-}},\qquad
b=A_{+}+A_{-}+2\sqrt{A_{+}A_{-}}.
\end{equation}
Similar expressions are valid also in terms of $B$'s and $C$'s.
Namely, the quantities \eqref{ABC} satisfy the
multiplicative self-consistency relations
\begin{equation}\label{mult}
A_{+}A_{-}=B_{+}B_{-}=C_{+}C_{-},
\end{equation}
together with the additive ones
\begin{equation}\label{addv}
A_{+}+A_{-}
=B_{+}+B_{-}-Q
=R-C_{+}-C_{-}.
\end{equation}
The first equation in \eqref{eqRII} now reads
\begin{equation}\label{RIIfirst}
\alpha\frac{C_{-}A_{+}}{C_{+}B_{-}}=1,
\end{equation}
and the second equation can be written in either of two forms
\begin{equation}\label{RIIsecond}
A_{\pm}+B_{\pm}+2C_{\pm}
=N_{\pm},
\end{equation}
where
\begin{equation}
N_{+}:= R+Q+1,\qquad
N_{-}:=R-1.
\end{equation}
Obviously, \eqref{mult}, \eqref{addv}, \eqref{RIIfirst}, and
\eqref{RIIsecond} constitute a system of six equations for the six
unknowns $A_\pm$, $B_\pm$, and $C_\pm$.

Before turning to solution of this system, it is useful to mention
that starting from one form of equation \eqref{RIIsecond},
say, using the `$+$' sign, the other form, with the `$-$' sign, can
be directly found from \eqref{addv}.  Indeed, rewriting \eqref{addv}
as
\begin{equation}\label{neweqs}
\begin{split}
A_{+}+A_{-}+C_{+}+C_{-}&=R,
\\
B_{+}+B_{-}+C_{+}+C_{-}&=R+Q,
\end{split}
\end{equation}
and summing these equations, it is clear that the two forms in  equation
\eqref{RIIsecond} are equivalent.

Dividing the two equations in \eqref{RIIsecond} by
$\sqrt{A_{\pm}B_{\pm}}$, respectively, and using the relation
$A_{+}/B_{+}=B_{-}/A_{-}$, which follows from the first equality in
\eqref{mult}, a comparison of the resulting relations yields
\begin{equation}
\frac{N_{+}-2C_{+}}{\sqrt{A_{+}B_{+}}}
=\frac{N_{-}-2C_{-}}{\sqrt{A_{-}B_{-}}}.
\end{equation}
Therefore, using \eqref{mult}, we can readily express $A$'s in
terms of $B$'s and $C$'s, as follows:
\begin{equation}\label{Aviaz}
A_{\pm}=B_{\mp}\frac{z_\pm}{z_\mp},
\end{equation}
where we have used the notation
\begin{equation}
z_\pm:=N_{\pm}-2C_{\pm}.
\end{equation}
We may now use these expressions in \eqref{neweqs} to eliminate the $B$'s.

Using \eqref{Aviaz}, and writing $C$'s in terms of $z$'s, from
\eqref{neweqs} we have
\begin{equation}
\begin{split}
B_{+}\frac{z_{-}}{z_{+}}+B_{-}\frac{z_{+}}{z_{-}}
&=\frac{z_{+}+z_{-}}{2}-\frac{Q}{2},
\\
B_{+}+B_{-}&=\frac{z_{+}+z_{-}}{2}+\frac{Q}{2}.
\end{split}
\end{equation}
Solving for $B$'s, we obtain
\begin{equation}\label{Bviaz}
B_{\pm}=\left(1\pm\frac{Q}{z_{+}-z_{-}}\right)\frac{z_{\pm}}{2}.
\end{equation}
Recalling that $B_{+}B_{-}=C_{+}C_{-}$, see \eqref{mult}, we have
\begin{equation}
\left(1-\frac{Q^2}{(z_{+}-z_{-})^2}\right)z_{+}z_{-}
=(N_{+}-z_{+})(N_{-}-z_{-}).
\end{equation}
Simplifying the $z_{+}z_{-}$ term standing in both sides, we
obtain the cubic equation:
\begin{equation}\label{curve}
Q^2 z_{+}z_{-}=
(N_{-}z_{+}+N_{+}z_{-}-N_{+}N_{+})(z_{+}-z_{-})^2.
\end{equation}
Note that it contains only the third and second order terms in $z$'s,
thus describing a rational curve. This means that $z$'s can be
expressed as a rational function of some parameter along the curve.

To obtain this parametrization, we set $z_{+}=tz_{-}$
in \eqref{curve} and consider $z_{-}$ as a
function of the parameter $t$. Then we get
\begin{equation}
z_{-}=\frac{1}{N_{-}t+N_{+}}\left[\frac{Q^2t}{(t-1)^2}+
N_{+}N_{-}\right],\qquad
z_{+}=tz_{-}.
\end{equation}
Note that we need just that portion of this curve for which the $C$'s
are positive.  It can be easily seen that this corresponds to values
of the parameter $t$ in the interval $[N_{+}/N_{-},\infty)$. It is
useful to introduce a new parameter $\eta\in [0,1]$, which parametrizes
this portion of the curve as follows:
\begin{equation}
t=\frac{N_{+}}{N_{-}}\frac{1+\eta}{1-\eta}.
\end{equation}
In terms of the new parameter, $z$'s read
\begin{equation}\label{zeta}
z_{\pm}
=N_{\pm}\frac{1\pm\eta}{2}
\left(\frac{Q^2(1-\eta)(1+\eta)}{\left(N_{+}(1+\eta)
-N_{-}(1-\eta)\right)^2}+1\right).
\end{equation}

As a result, recalling that $C_\pm=(E_\pm-z_\pm)/2$, using expressions
for $B$'s \eqref{Bviaz}, and those for $A$'s \eqref{Aviaz}, we obtain
that the parametrization \eqref{zeta} implies nice factorized
expressions:
\begin{equation}\label{ABCs}
\begin{split}
C_{+}
&=N_{+}
\frac{(1-\eta)(1+R\eta)\big(1+Q+(R+Q)\eta\big)}{\big(2+Q+(2R+Q)\eta\big)^2},
\\
C_{-}
&=N_{-}
\frac{(1+\eta)(1+Q+R\eta)\big(1+(R+Q)\eta\big)}{\big(2+Q+(2R+Q)\eta\big)^2},
\\
B_{+}
&=N_{+}
\frac{(1+\eta)(1+Q+R\eta)\big(1+Q+(R+Q)\eta\big)}{\big(2+Q+(2R+Q)\eta\big)^2},
\\
B_{-}
&=N_{-}
\frac{(1-\eta)(1+R\eta)\big(1+(R+Q)\eta\big)}{\big(2+Q+(2R+Q)\eta\big)^2},
\\
A_{+}
&=N_{+}
\frac{(1+\eta)(1+R\eta)\big(1+(R+Q)\eta\big)}{\big(2+Q+(2R+Q)\eta\big)^2},
\\
A_{-}
&=N_{-}
\frac{(1-\eta)(1+Q+R\eta)\big(1+Q+(R+Q)\eta\big)}{\big(2+Q+(2R+Q)\eta\big)^2}.
\end{split}
\end{equation}
These expressions solve equations \eqref{mult}, \eqref{addv}, and
\eqref{RIIsecond}, and moreover they correspond to the solution in
which all $A$'s, $B$'s, and $C$'s are positive, provided that
$\eta\in[0,1]$.

Note that till now equation \eqref{RIIfirst} has never been used yet.
To obtain the solution of the initial end-points problem, we exploit
this last equation.  Using \eqref{ABCs}, we have
\begin{equation}\label{alpha-eq}
\alpha \frac{(1+\eta)^2(1+Q+R\eta)\big(1+(R+Q)\eta\big)}
{(1-\eta)^2(1+R\eta)\big(1+Q+(R+Q)\eta\big)}=1.
\end{equation}
This is a quartic equation, and we need only that root taking values
in the interval $[0,1]$. It can be verified that such a
root $\eta=\eta(R,Q;\alpha)$ always exists, provided that the
parameters $R$, $Q$, and $\alpha$ correspond to Regime II. An explicit
expression for this root is not crucial for what follows; it is
discussed in Appendix A.

\subsection{Regime II: The free energy}

In order to extract the first moment of the equilibrium measure in
Regime IIA and Regime IIB, we evaluate the term $O(1/z^2)$ in the
$z\to\infty$ expansion of the resolvent, see \eqref{WIIA} and
\eqref{WIIB}, respectively. In both regimes we obtain
\begin{equation}
E=\frac{2+Q}{8}(a+b)+\frac{Q^2}{4}-\frac{Q}{4}\sqrt{(a+Q)(b+Q)}
+\frac{R}{2}\sqrt{(R-a)(R-b)}.
\end{equation}
Expressing $E$ in terms of quantities \eqref{ABC}, we have
\begin{equation}
E=\frac{Q^2}{4}+\frac{2+Q}{4}(A_{+}+A_{-})-\frac{Q}{4}(B_{+}-B_{-})
+\frac{R}{2}(C_{+}-C_{-}).
\end{equation}
To obtain the function $\Phi(R,Q)$, we write the integral
in \eqref{intE} as
\begin{equation}
\Phi(R,Q)=\int E \frac{\partial\log\alpha}{\partial \eta} \, \rmd\eta+C(R,Q),
\end{equation}
where the functions $E=E(\eta)$ and $\alpha=\alpha(\eta)$
follow from \eqref{ABCs} and \eqref{alpha-eq}, respectively.

Evaluating the integral, we find that
\begin{equation}\label{PhiRQ}
\Phi(R,Q)=\Omega(R,Q)+C(R,Q),
\end{equation}
where the function $\Omega(R,Q)=\Omega(R,Q;\eta)$ is given as a linear
combination of seven logarithms:
\begin{multline}\label{Omega}
\Omega(R,Q)
=\left(\frac{1}{2}+R-\frac{R^2}{2}\right)\log(1-\eta)
+\left(\frac{1}{2}-R+Q-\frac{(R+Q)^2}{2}\right)\log(1+\eta)
\\
+\left(\frac{1}{2}+R\right)\log(1+R\eta)
+\left(\frac{1}{2}-R+Q-RQ\right)\log(1+Q+R\eta)
\\
+ \left(\frac{1}{2}-R\right)\log\big(1+(R+Q)\eta\big)
\\
+\left(\frac{1}{2}+R+Q+RQ+Q^2\right)\log\big(1+Q+(R+Q)\eta\big)
\\
-\frac{(2+Q)^2}{2}\log\big(2+Q+(2R+Q)\eta\big).
\end{multline}
To find the integration constant $C(R,Q)$, we recall that in our
preliminary discussion we obtained the condition \eqref{Phiat1}.  To
employ this condition, we note that from \eqref{alpha-eq} it follows
that $\eta\to0$ as $\alpha\to 1$, hence the constant $C(R,Q)$ can be
inferred by equating \eqref{PhiRQ} at $\eta=0$ with the expression
\eqref{Phiat1}, that gives
\begin{multline}\label{CRQ}
C(R,Q)=-\frac{R^2}{2}\log R+\frac{(R-1)^2}{2}\log (R-1)
-\frac{(R+Q)^2}{2} \log (R+Q)
\\
-\frac{(1+Q)^2}{2}\log(1+Q)
+\frac{(R+Q+1)^2}{2}\log(R+Q+1).
\end{multline}
Expressions \eqref{PhiRQ}, \eqref{Omega}, and \eqref{CRQ} provide a
solution of the free energy problem of the discrete Coulomb gas in
Regime II.

From \eqref{sigmaPhi} and \eqref{RQdef}, it follows that in Regime II
the function $\varphi(x,y)$ is given by the following expression:
\begin{equation}\label{chi-chi}
\varphi(x,y)=\chi(x,y;\eta)-\chi_0(x,y),
\qquad (x,y)\in\mathcal{D}_{\mathrm{II}},
\end{equation}
where
\begin{multline}\label{sigmalong}
\chi(x,y;\eta)=-xy\log(1-\alpha)+\frac{y^2}{2}\log\alpha
\\
+\left(\frac{x^2}{2}+xy-\frac{y^2}{2}-x-y+\frac{1}{2}\right)\log(1-\eta)
+\left(y^2-2xy+\frac{1}{2}\right)\log(1+\eta)
\\
-\left(\frac{y}{2}-x+1\right)y\log\big[y+(1-x)\eta\big]
-\left(x^2+xy-\frac{y^2}{2}-x\right)\log\big[x+(1-x)\eta\big]
\\
-\left(\frac{y}{2}+x-1\right)y\log\big[y+(1-y)\eta\big]
-\left(\frac{y^2}{2}-xy+x\right)\log\big[x+(1-y)\eta\big],
\\
+\frac{(x+y)^2}{2}\log\big[x+y+(2-x-y)\eta\big],
\end{multline}
and the function $\chi_0(x,y)$, independent of $\eta$, reads
\begin{multline}\label{chi0}
\chi_0(x,y)=
-\frac{x^2}{2}\log x-\frac{(1-x)^2}{2}\log(1-x)
-\frac{y^2}{2}\log y-\frac{(1-y)^2}{2}\log(1-y)
\\
+\frac{(1-x-y)^2}{2}\log(1-x-y).
\end{multline}

Taking into account that the equation for the parameter $\eta$ in
terms of variables $x$ and $y$ reads
\begin{equation}\label{etaeq}
\alpha\frac{(1+\eta)^2\big[x+(1-x)\eta\big]\big[y+(1-y)\eta\big]}
{(1-\eta)^2\big[y+(1-x)\eta\big]\big[x+(1-y)\eta\big]}=1
\end{equation}
and also using that
\begin{equation}
1-\alpha=\frac{\eta\big[x+y+(2-x-y)\eta\big]^2}
{(1+\eta)^2\big[x+(1-x)\eta\big]\big[y+(1-y)\eta\big]},
\end{equation}
we may express $\alpha$ in terms of $\eta$ in \eqref{sigmalong},
thus obtaining for $\chi(x,y;\eta)$ the following expression:
\begin{multline}\label{chieta}
\chi(x,y;\eta)
=-xy\log\eta+\frac{(1-x-y)^2}{2}\log(1-\eta)+\frac{1}{2}\log(1+\eta)
\\
-(1-x)y\log\big[y+(1-x)\eta\big]
-x(1-y)\log\big[x+(1-y)\eta\big]
\\
+x(1-x)\log\big[x+(1-x)\eta\big]
+y(1-y)\log\big[y+(1-y)\eta\big]
\\
+\frac{(x-y)^2}{2}\log\big[x+y+(2-x-y)\eta\big].
\end{multline}

We conclude by stating a remarkable property of representation
\eqref{chi-chi}.
\begin{proposition} Let:
\begin{equation}
h=h(x,y):=\sqrt{\frac{xy}{(1-x)(1-y)}}.
\end{equation}
Then  the following holds:
\begin{equation}\label{chi0chih}
\chi_0(x,y)=\chi(x,y;h).
\end{equation}
\end{proposition}
Verification is fairly straightforward.  Representation
\eqref{chi-chi} together with \eqref{chieta} and \eqref{chi0chih}
immediately implies formula \eqref{mainresult}, which is thus proven.

The proofs of Propositions \ref{propa3} and \ref{propt3} are based on
specific properties of the function $\chi(x,y;\eta)$, namely that its
first and second derivative with respect to $\eta$ vanish at $\eta=h$.
Further details and explicit calculations are given in Appendix B.


\appendix
\section{End-point equations in Regime II: A direct treatment}

We discuss here a rather direct approach to treat the end-point
equations \eqref{eqRIIA}. Using notations \eqref{ABC}, we note that
these equations can be written using only $A_{+}$, $B_{+}$, and
$C_{+}$, or only $A_{-}$, $B_{-}$, and $C_{-}$,
\begin{equation}
\begin{split}
\frac{C_{\pm}}{\sqrt{A_{\pm}B_{\pm}}}
&=(\sqrt{\alpha})^{\pm1},
\\
A_{\pm}+B_{\pm}+2C_{\pm}
&=N_\pm.
\end{split}
\end{equation}
Introduce parameter $\gamma$ by
\begin{equation}\label{gamma}
\gamma:=\frac{B_{-}}{A_{-}}=\frac{A_{+}}{B_{+}}.
\end{equation}
From the first and second way of writing of the end-point equations, we
have, respectively,
\begin{equation}
A_{+}=\frac{\gamma N_{+}}{1+\gamma+2\sqrt{\alpha\gamma}},\qquad
A_{-}=\frac{\sqrt{\alpha}N_{-}}{(1+\gamma)\sqrt{\alpha}+2\sqrt{\gamma}}.
\end{equation}
From the definition of the quantities $A_\pm$  we have
\begin{equation}
a=A_{+}+A_{-}-2\sqrt{A_{+}A_{-}},\qquad
b=A_{+}+A_{-}+2\sqrt{A_{+}A_{-}},
\end{equation}
and
\begin{equation}
\sqrt{a+Q}=\sqrt{\frac{A_{+}}{\gamma}}-\sqrt{\gamma A_{-}},\qquad
\sqrt{b+Q}=\sqrt{\frac{A_{+}}{\gamma}}+\sqrt{\gamma A_{-}}.
\end{equation}
The expressions for the end-points $a$ and $b$ which follow from these
two ways of writing are compatible with each other, provided that
\begin{equation}
Q+A_{+}+A_{-}=\frac{A_{+}}{\gamma}+\gamma A_{-}.
\end{equation}
Explicitly, this is a an equation for the parameter $\gamma$,
\begin{equation}
Q+(1-\gamma)
\frac{\sqrt{\alpha}N_{-}}{(1+\gamma)\sqrt{\alpha}+2\sqrt{\gamma}}=
(1-\gamma)\frac{N_{+}}{1+\gamma+2\sqrt{\alpha\gamma}}.
\end{equation}
Dividing by the expression standing in the right-hand side,
\begin{equation}
\frac{Q}{N_{+}} \frac{1+\gamma+2\sqrt{\alpha\gamma}}{1-\gamma}
+\frac{\sqrt{\alpha}N_{-}}{N_{+}}
\frac{1+\gamma+2\sqrt{\alpha\gamma}}{(1+\gamma)\sqrt{\alpha}+2\sqrt{\gamma}}=1,
\end{equation}
and introducing a new parameter $\omega$ such that
\begin{equation}\label{cosomega}
\cos\omega=
\frac{(1+\gamma)\sqrt{\alpha}+2\sqrt{\gamma}}{1+\gamma+2\sqrt{\alpha\gamma}},
\qquad
\sin\omega= \frac{\sqrt{1-\alpha}(1-\gamma)}{1+\gamma+2\sqrt{\alpha\gamma}},
\end{equation}
we may rewrite the equation in the form
\begin{equation}\label{AB1}
\frac{A}{\cos\omega}+\frac{B}{\sin\omega}
=1,
\end{equation}
where
\begin{equation}\label{ABcoef}
A=\frac{\sqrt{\alpha} N_{-}}{N_{+}},\quad
B=\frac{\sqrt{1-\alpha}\,Q}{N_{+}}.
\end{equation}
It is interesting to note that the three parameters ($\alpha$, $R$,
and $Q$) in fact enters the equation only through their two
combinations, $A$ and $B$.

It is possible to find the roots of \eqref{AB1}, by noting that
it can be written as a quartic equation
\begin{equation}
x^4-2Bx^3-(1-A^2-B^2)x^2+2Bx-B^2=0,\qquad
x=\sin\omega.
\end{equation}
The roots of this equation are
\begin{multline}\label{roots}
x=\frac{B}{2}+\frac{1}{2}\Bigg\{
\sqrt{B^2+2S\left(1-\cos\frac{\theta}{3}\right)}
\pm\sqrt{B^2+2S\left(1-\cos\frac{\theta+2\pi}{3}\right)}
\\
\pm\sqrt{B^2+2S\left(1-\cos\frac{\theta-2\pi}{3}\right)}
\Bigg\},
\end{multline}
where
\begin{equation}
\theta\in[0,\pi),\qquad
\theta=\arccos\left(1-\frac{2A^2B^2}{S^3}\right),
\qquad S=\frac{1-A^2-B^2}{3}>0,
\end{equation}
and the signs are to be chosen independently.  The root which lies in
the interval $[0,1]$ corresponds to the first and second signs in
\eqref{roots} being `$+$' and `$-$', respectively, as it can be
checked by expanding the expression in \eqref{roots} in power series,
say, in $B$, and noting that the required root has the behaviour
$x\sim (1-A)^{-1}B$, as $B\to 0$.

To find a relation between $\eta$ and parameter $\varphi$, we first
express parameter $\gamma$, \eqref{gamma}, in terms of $\eta$ using
\eqref{ABCs},
\begin{equation}
\gamma=\frac{(1+R\eta)\big(1+(R+Q)\eta\big)}{(1+Q+R\eta)(1+Q+(R+Q)\eta)}.
\end{equation}
Using \eqref{alpha-eq} we also have
\begin{equation}
\sqrt{\alpha\gamma}=\frac{(1-\eta)(1+R\eta)}{(1+\eta)(1+Q+R\eta)},
\qquad\sqrt{\frac{\gamma}{\alpha}}
=\frac{(1+\eta)(1-R\eta)}{(1+(R+Q)\eta)(1+Q+(R+Q)\eta)},
\end{equation}
and substituting in the expression for, say, $\cos\omega$, see
\eqref{cosomega}, we obtain
\begin{equation}
\cos\omega=\sqrt{\alpha}\frac{1+\eta}{1-\eta}.
\end{equation}
Consequently,
\begin{equation}\label{eta-omega}
\eta=\frac{\cos\omega-\sqrt{\alpha}}{\cos\omega+\sqrt{\alpha}}.
\end{equation}
An expression involving $\sin\omega$ can be given by taking into
account \eqref{AB1}. Thus \eqref{roots}, with the indicated choice of
the signs, provides an explicit expression for $\eta$.

\section{Phase transition at the Arctic ellipse}

Here we outline some calculations related to the third-order phase
transition at the curve $\mathcal{A}$ separating the regions
$\mathcal{D}_\mathrm{I}$ and $\mathcal{D}_\mathrm{II}$, see
Fig.~\ref{fig-ArcticEllipse}.

We first mention a particular parametrization of the points of
$\mathcal{A}$ which appears useful in these calculations.  Recalling
that $\mathcal{A}$ is a portion of the ellipse
\begin{equation}
\frac{(1-x-y)^2}{\alpha}+
\frac{(x-y)^2}{1-\alpha}=0,
\end{equation}
one can introduce an angle parameter $\phi$, such that
\begin{equation}
1-x-y=\sqrt{\alpha} \sin\phi,\quad x-y=\sqrt{1-\alpha}\cos\phi.
\end{equation}
If we further set $\alpha=\sin^2\lambda$, $\lambda\in(0,\pi)$, then we
have
\begin{equation}\label{x0y0}
x=\cos^2\left(\frac{\phi+\lambda}{2}\right),\quad
y=\sin^2\left(\frac{\phi-\lambda}{2}\right),
\qquad (x,y)\in\mathcal{A},\quad \phi\in[\lambda,\pi/2].
\end{equation}

To illustrate the convenience of this parametrization, let us consider
our quartic equation in the case of $(x,y)\in\mathcal{A}$. We take it
in its $\omega$-parametrized form \eqref{AB1}. Recalling relations
\eqref{RQdef}, for the coefficients $A$ and $B$, see \eqref{ABcoef},
we have $A=\sqrt{\alpha}(1-x-y)$ and $B=\sqrt{1-\alpha}(x-y)$, and so
\eqref{AB1} in this case reads
\begin{equation}
\frac{\sin^2\lambda\sin\phi}{\cos\omega}
+\frac{\cos^2\lambda\cos\phi}{\sin\omega}=1.
\end{equation}
The root we need is simply $\omega=\pi/2-\phi$ (note that
$\phi\in[\lambda,\pi/2]$ and $\lambda\in(0,\pi/2)$, and hence
$\omega\in[0,\pi/2)$, as required). Substituting the obtained
expression for the root in \eqref{eta-omega}, we get
\begin{equation}
\eta\big|_{(x,y)\in\mathcal{A}}
=\frac{\sin\phi-\sin\lambda}{\sin\phi+\sin\lambda}
=\cot\left(\frac{\phi+\lambda}{2}\right)
\tan\left(\frac{\phi-\lambda}{2}\right)
=\sqrt{\frac{xy}{(1-x)(1-y)}}.
\end{equation}
In the last expression one can recognize the quantity $h$, see
\eqref{hxy}, and hence its interpretation as the value of $\eta$ on
the Arctic ellipse.

Let us now turn to study the properties of the function
$\varphi(x,y)=\varphi(x,y;\alpha)$ in the vicinity of $\mathcal{A}$.
In particular, we want to verify that
\begin{equation}
\varphi(x,y)\sim C (\alpha-\ac)^3,\qquad  \alpha\to\ac^+.
\end{equation}
Since $\varphi(x,y)=\chi(x,y;\eta)-\chi(x,y;h)$ where the function
$\chi(x,y;\eta)$ is given by \eqref{chieta}, and since the dependence
on $\alpha$ enters only through $\eta$, the problem reduces to
studying properties of $\chi(x,y;\eta)$ as a function $\eta$. By a
direct calculation it can be verified that
\begin{equation}\label{dchi0}
\partial_\eta^k\chi(x,y;\eta)\big|_{\eta=h}=0,\qquad k=1,2.
\end{equation}
At the same time the third-order derivative does not vanish, that gives
a non-zero value of $C=C(x,y)$, which can computed by the formula
\begin{equation}
C=\partial_\alpha^3\varphi\big|_{\alpha=\ac}=
\frac{1}{\ac^3}
\left[\frac{1}{(\partial_\eta \log\alpha)^3}
\partial_\eta^3 \chi(x,y;\eta)\right]\bigg|_{\eta=h},
\end{equation}
where the function $\alpha=\alpha(\eta)$ is determined by \eqref{etaeq}.
By a straightforward calculation we obtain
\begin{multline}
\partial_\eta \log\alpha\big|_{\eta=h} =
-\frac{2}{1-h}-\frac{2}{1+h}+\frac{1-x}{y+(1-x)h}+\frac{1-y}{x+(1-y)h}
\\
-\frac{1-x}{x+(1-x)h}-\frac{1-y}{y+(1-y)h}
=
-4 \frac{(1-x)(1-y)}{1-x-y}.
\end{multline}
However, in the case of $\partial_\eta^3 \chi(x,y;\eta)$ the corresponding
calculation appears to be much more involved; the parametrization
\eqref{x0y0} turns out to be very convenient (in particular, when
resorting to computer symbolic calculations), providing the result in
a nice factorized form:
\begin{equation}
\partial_\eta^3 \chi(x,y;\eta)\big|_{\eta=h}=
-\frac{(\sin\phi-\sin\lambda)(\sin\phi+\sin\lambda)^7}
{8\sin\lambda\cos{\!}^4\lambda\sin{\!}^3\phi}.
\end{equation}
Taking into account that $\partial_\eta \log\alpha\big|_{\eta=h}
=-(\sin\phi+\sin\lambda)^2/\sin\lambda\sin\phi$ and
$\ac=\sin^2\lambda$, we finally obtain
\begin{equation}
C=\frac{2\sin(\phi-\lambda)\sin(\phi+\lambda)}{\sin^4 2\lambda}=
\frac{\sqrt{x(1-x)y(1-y)}}{2\ac^2(1-\ac)^2},
\end{equation}
and we recall that $\ac$ as a function of $x$ and $y$ is given by
\eqref{avw}.

Similarly, the third-order phase transition can be regarded as a
property of the function $\varphi(x,y)=\varphi(x,y;\alpha)$ with
respect to the variables $x$ and $y$, at fixed $\alpha$. Namely, the
function $\varphi(x,y)=\varphi(x,y;\alpha)$ has a discontinuity in its
third normal derivative at the curve $\mathcal{A}$, with vanishing
first and second normal derivative. Let $t$ be the variable along the
normal of the curve $\mathcal{A}$ at the point $(x,y)$, then
\begin{equation}\label{dtvarphi0}
\partial_t^k\varphi(x,y)|_{(x,y)\in\mathcal{A}}=0,\qquad k=1,2.
\end{equation}
Since $\varphi(x,y)=\chi(x,y;\eta)-\chi(x,y;h)$, in computing the
derivatives with respect to $t$ one has to take into account only
those terms which involve derivatives of the functions
$\chi(x,y;\eta)$ and $\chi(x,y;h)$ with respect to their parameters,
$\eta$ and $h$, respectively. Hence the property \eqref{dtvarphi0} is
just a consequence of the analogous property for the function
$\chi(x,y;\eta)$, see \eqref{dchi0}.  For the third derivative we have
\begin{equation}
\partial_t^3\varphi(x,y)|_{(x,y)\in\mathcal{A}}=
\left[(\partial_t\eta|_{\eta=h})^3-(\partial_t h)^3\right]
\partial_\eta^3 \chi(x,y;\eta)\big|_{\eta=h}.
\end{equation}
Since $\partial_t\eta|_{\eta=h}$ in general is not equal to
$\partial_t h$, the third derivative is non-vanishing; its explicit
expression can be obtained along the same lines as for the quantity
$C$ above.

\bibliography{rer_bib}
\end{document}